\documentclass[%
aip,apl,
 amsmath,amssymb,
 reprint,%
]{revtex4-1}

\usepackage{graphicx}
\usepackage{dcolumn}
\usepackage{bm}

\usepackage[utf8]{inputenc}
\usepackage[T1]{fontenc}
\usepackage{mathptmx}
\usepackage{etoolbox}

\usepackage{hyperref}
\hypersetup{
colorlinks=true,
urlcolor= blue,
citecolor=blue,
linkcolor= blue,
bookmarks=true,
bookmarksopen=false,
}

\begin{document}


\title{Epitaxial growth and scanning tunneling microscopy of LiV$_2$O$_4$ thin films on SrTiO$_3$(111)} 



\author{T. F. Schweizer}
\affiliation{Max Planck Institute for Solid State Research, 70569 Stuttgart, Germany}

\author{U. Niemann}
\affiliation{Max Planck Institute for Solid State Research, 70569 Stuttgart, Germany}

\author{X. Que}
\affiliation{Max Planck Institute for Solid State Research, 70569 Stuttgart, Germany}

\author{Q. He}
\affiliation{Max Planck Institute for Solid State Research, 70569 Stuttgart, Germany}

\author{L. Zhou}
\affiliation{Max Planck Institute for Solid State Research, 70569 Stuttgart, Germany}

\author{M. Kim}
\affiliation{Max Planck Institute for Solid State Research, 70569 Stuttgart, Germany}

\author{H. Takagi}
\affiliation{Max Planck Institute for Solid State Research, 70569 Stuttgart, Germany}
\affiliation{Institute for Functional Matter and Quantum Technologies, University of Stuttgart, 70569 Stuttgart, Germany}
\affiliation{Department of Physics, The University of Tokyo, 113-0033 Tokyo, Japan}

\author{D. Huang}
\email[]{D.Huang@fkf.mpg.de}
\affiliation{Max Planck Institute for Solid State Research, 70569 Stuttgart, Germany}


\date{\today}

\begin{abstract}
LiV$_2$O$_4$ is a mixed-valent spinel oxide and one of a few transition-metal compounds to host a heavy fermion phase at low temperatures. While numerous experimental studies have attempted to elucidate how its 3$d$ electrons undergo giant mass renormalization, spectroscopic probes that may provide crucial hints, such as scanning tunneling microscopy (STM), remain to be applied. A prerequisite is atomically flat and pristine surfaces, which, in the case of LiV$_2$O$_4$, are difficult to obtain by cleavage of small, three-dimensional crystals. We report the epitaxial growth of LiV$_2$O$_4$ thin films with bulklike properties on SrTiO$_3$(111) via pulsed laser deposition and stable STM imaging of the LiV$_2$O$_4$(111) surface. The as-grown films were transferred \textit{ex situ} to a room-temperature STM, where subsequent annealing with optional sputtering in ultrahigh vacuum enabled compact islands with smooth surfaces and a hexagonal 1$\times$1 atomic lattice to be resolved. Our STM measurements provide insights into growth mechanisms of LiV$_2$O$_4$ on SrTiO$_3$(111), as well as demonstrate the feasibility of performing surface-sensitive measurements of this heavy fermion compound.
\end{abstract}


\maketitle 


The mixed-valent spinel oxide LiV$_2$O$_4$ has drawn attention over the years as the first discovered example of a heavy fermion system with 3$d$ electrons, instead of 4$f$ or 5$f$ electrons \cite{Kondo_PRL_1997, Urano_PRL_2000}. A key feature of this compound is the geometric frustration experienced by both its spins and charges. The magnetic V atoms occupy a pyrochlore sublattice with corner-sharing tetrahedra, which engenders spin frustration in the presence of nearest-neighbor antiferromagnetic exchange \cite{Kessler_JCP_1971}. The nominal V$^{3.5+}$ ions are further hindered from charge disproportionation into a 1:1 population of V$^{3+}$ and V$^{4+}$ species. In order to minimize the short-range Coulomb energy, the number of adjacent V$^{3+}$-V$^{4+}$ pairs should be maximized, in analogy to the Ising model. However, a macroscopic number of degenerate configurations exists on the pyrochlore sublattice, leading to charge frustration \cite{Anderson_PR_1956}. While in many other mixed-valent spinel oxides \cite{Verwey_Nature_1939, Matsuno_JPSJ_2001, Yamada_MRB_1995, Okamoto_PRL_2008, Browne_IC_2018}, the frustration is eventually relieved by structural distortions coupled with spin, charge, and/or orbital orders, LiV$_2$O$_4$ remains cubic with equivalent V sites down to at least 4 K \cite{Chmaissem_PRL_1997}, with no signs of ordering.

At low temperatures, LiV$_2$O$_4$ exhibits a metallic phase that satisfies many hallmarks of the heavy fermion state seen in Kondo-lattice systems with $f$ electrons. The linear coefficient $\gamma$ extracted from the electronic heat capacity has a sizeable value of 0.42 J/(mol K$^2$) at 1 K \cite{Kondo_PRL_1997}, which implies a 25-fold enhancement of effective mass compared to the value predicted by first-principles calculations with the local density approximation \cite{Matsuno_PRB_1999}. The spin susceptibility $\chi$ is also large and nearly constant at low temperatures. Together, $\chi$ and $\gamma$ yield a Wilson ratio of 1.7 \cite{Kondo_PRL_1997}, which is close to the value of one for free electrons. Resistivity measurements of single crystals showed on cooling a downturn at $T^* =$ 20--30 K from $T$-linear-like behavior,  followed by a $T^2$-like temperature dependence characteristic of a Fermi liquid below 2 K \cite{Urano_PRL_2000}. From a fit of the resistivity to $\rho = \rho_0 + A T^2$, Urano \textit{et al.}~\cite{Urano_PRL_2000} estimated the Kadowaki-Woods ratio to be $A/\gamma \approx 10^{-5}$ ($\mu\Omega$ cm/K$^2$) / [mJ/(mol K$^2$)]$^2$, which is comparable to that of other heavy fermion compounds with $f$ electrons. Despite these similarities, it is unclear how the 3$d$ electrons in LiV$_2$O$_4$ can be mapped onto a Kondo lattice, or whether a distinct mechanism of mass enhancement is at play, and such questions have inspired a large number of experimental studies on bulk samples \cite{Kondo_PRL_1997, Urano_PRL_2000, Shimoyamada_PRL_2006, Jonsson_PRL_2007, Shimizu_NatComm_2012, Browne_PRM_2020} and theoretical models \cite{Anisimov_PRL_1999, Fulde_EPL_2001, Burdin_PRB_2002, Fujimoto_PRB_2002, Hopkinson_PRL_2002, Yamashita_PRB_2003, Arita_PRL_2007, Hattori_PRB_2009}.

Thin films of LiV$_2$O$_4$ grown by pulsed laser deposition (PLD) represent a nascent avenue to derive fresh insights into the heavy fermion enigma \cite{Yajima_PRB_2021, Niemann_arXiv_2022}. Thin films are especially amenable to charge and lattice engineering, and both electrochemical Li intercalation \cite{Yajima_PRB_2021} and epitaxial strain \cite{Niemann_arXiv_2022} have been shown to induce metal-to-insulator transitions in LiV$_2$O$_4$ films. The latter is especially revealing, as tensile strain in the (001) plane explicitly breaks the degeneracy of the V pyrochlore sublattice. The resulting emergence of a Verwey-type charge order in strained, tetragonal films strongly implies that the heavy fermion phase in cubic LiV$_2$O$_4$ is a charge-frustrated phase, and that intense charge fluctuations underlie the giant mass renormalization of its 3$d$ electrons \cite{Niemann_arXiv_2022}. Thin films are also amenable to angle-resolved photoemission spectroscopy and scanning tunneling microscopy (STM), which can be used to determine which electronic band (or bands) becomes heavy at low temperatures. STM can also elucidate the surface structure of LiV$_2$O$_4$, which is useful for potential electrochemical applications \cite{Lu_AMI_2019}. STM requires atomically flat and pristine surfaces, which are challenging to achieve with the typical size and morphology of LiV$_2$O$_4$ crystals \cite{Das_PRB_2007}. With thin films grown on typical millimeter-size substrates, however, atomically resolved STM has been proven possible on a number of closely related spinel oxides, including Fe$_3$O$_4$(111)~\cite{Ritter_SS_1999}, Co$_3$O$_4$(111)~\cite{Meyer_JPCM_2008}, LiTi$_2$O$_4$(111)~\cite{Okada_NatComm_2017}, and Li$_4$Ti$_5$O$_{12}$(111)~\cite{Kitta_SS_2014}. These successes motivate our present study.

\begin{figure*}
\includegraphics{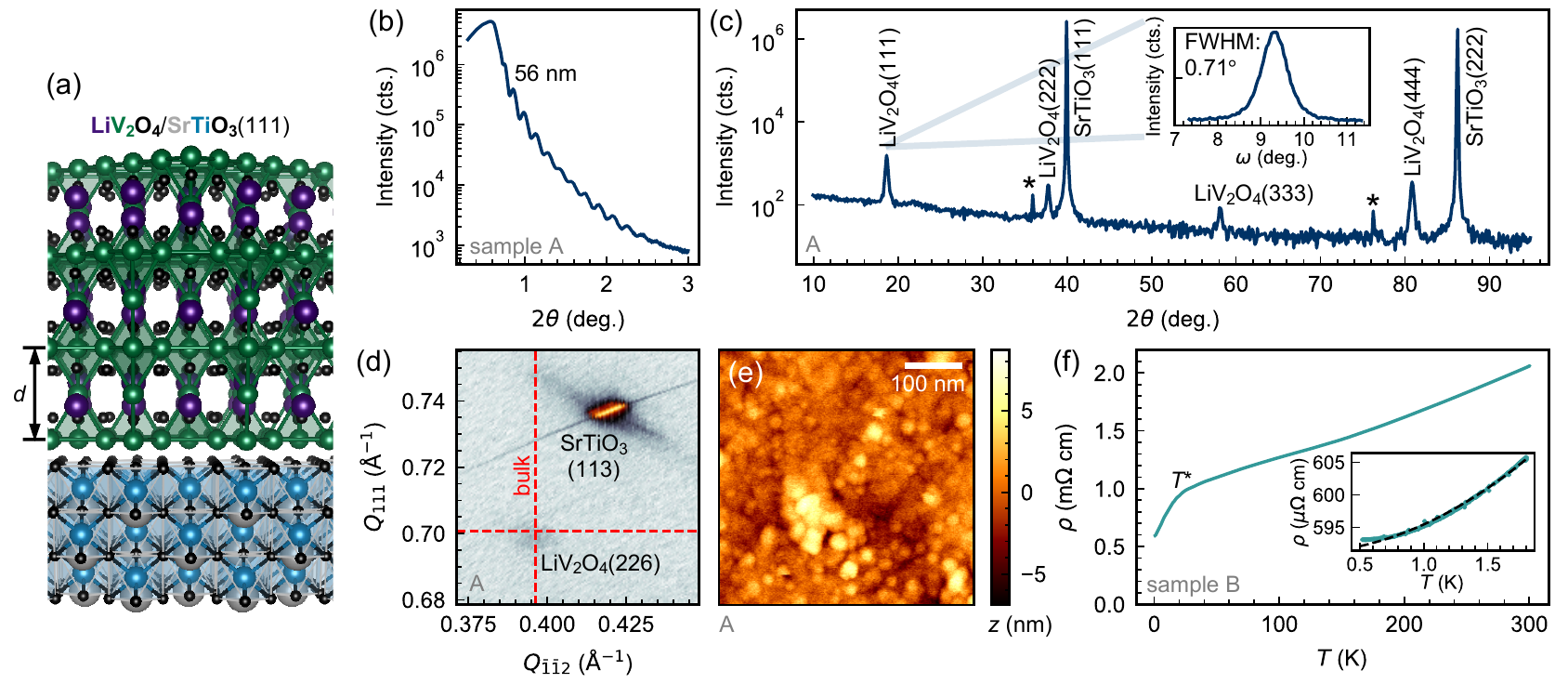}
\caption{\label{Fig1} (a) Schematic structure of a LiV$_2$O$_4$/SrTiO$_3$(111) film. The lattice spacing along the [111] direction is marked by $d$. (b) X-ray reflectivity oscillations of sample A. The estimated film thickness is 56 nm. (c) X-ray $\omega$-2$\theta$ scan. The asterisks mark spurious reflections of the dominant substrate peaks SrTiO$_3$(111) and SrTiO$_3$(222) due to an unfiltered Cu K$_{\beta}$ component of the Cu K$_{\alpha}$ source. Inset: rocking curve of the LiV$_2$O$_4$(111) peak, with a FWHM of 0.71$^{\circ}$. (d) X-ray reciprocal space mapping around the SrTiO$_3$(113) peak, revealing a LiV$_2$O$_4$(226) peak with relaxed, bulk-like values \cite{Chmaissem_PRL_1997}, as marked by the dotted lines. $Q_{\bar{1}\bar{1}2}$ and $Q_{111}$ denote the distance in reciprocal space from the origin along the [$\bar{1}$$\bar{1}$2] and [111] directions, respectively. (e) AFM topographic image of film. The root-mean-square roughness over the 500$\times$500 nm area is 1.7 nm. (f) Temperature-dependent resistivity of sample B, which shows metallic behavior and a characteristic downturn at $T^*$. Inset: enlarged view between 0.5 and 1.8 K and a quadratic fit to the data. The deviation near 0.5 K may arise from residual impurities.}
\end{figure*}

Here, we demonstrate that atomically resolved STM is possible on the surface of LiV$_2$O$_4$ films grown on SrTiO$_3$(111) substrates via PLD. A key step to preparing flat and pristine surfaces, even after exposing the films to air, is a high-temperature anneal in ultrahigh-vacuum (UHV), with or without Ar sputtering, monitored by reflection high-energy electron diffraction (RHEED). STM topographic images reveal a terrain of compact, flat islands with structural defects such as irregular step heights, tilted planes, screws dislocations, and trenches. We resolved a hexagonal 1$\times$1 atomic lattice whose bright spots likely coincide with the location of V atoms.

PLD of LiV$_2$O$_4$ thin films was carried out following Ref.~\cite{Niemann_arXiv_2022} in a homebuilt chamber integrated with a KrF excimer laser of wavelength 248 nm (Coherent LPX PRO 210F). Substrates of insulating SrTiO$_3$(111) and conducting 0.5 wt\% Nb-SrTiO$_3$(111) were pretreated by annealing up to 1000$^{\circ}$C in atmosphere in order to obtain atomically flat terraces (see Fig.~S1 in the supplementary material). Our choice of SrTiO$_3$ was motivated by its widespread use in thin-film growth as a workhorse substrate, as well as the ability to tune its conductivity via Nb doping for different applications. We picked [111]-oriented substrates in order to induce [111]-oriented growth of LiV$_2$O$_4$, as the (111) termination of spinels have been suggested to possess the lowest surface energy \cite{Mishra_JAP_1977, Wen_PRB_2020}. For the target, we used a sintered pellet produced from mixing LiCO$_3$ and V$_2$O$_5$ with a molar ratio of 1.75:1. The excess nominal Li:V ratio of 1.75:2, compared to 1:2 for LiV$_2$O$_4$, was intended to compensate for the loss of light Li atoms upon laser ablation. The substrate temperature ranged from 520--660$^{\circ}$C (see Fig.~S2 in the supplementary material for additional information). Deposition took place in high vacuum conditions (base pressure 10$^{-7}$ mbar) with no additional supply of O$_2$ gas. We used a pulse repetition rate of 5 Hz and a laser fluence of approximately 1.5--1.7 J/cm$^2$. Atomic force microscopy (AFM) was performed with a Bruker system using PeakForce tapping. STM at room temperature was performed in a separate custom system (Unisoku). We used electrochemically etched W tips that were subsequently sharpened \textit{in situ} by electron-beam heating. STM topographies were acquired in constant-current imaging mode, and high scan speeds up to 80 nm/s were needed to overcome large thermal drifts at room temperature. Piezo calibration in the $z$ direction was performed on step edges of a Au(111)/mica sample (see Fig.~S3 in the supplementary material).

Figures \ref{Fig1}(a)--\ref{Fig1}(d) present x-ray diffraction of a LiV$_2$O$_4$ film deposited on insulating SrTiO$_3$(111) (sample A). We estimate the film thickness to be 56 nm, based on Kiessig oscillations in x-ray reflectivity measurements at grazing angles [Fig.~\ref{Fig1}(b)]. In the $\omega$-$2\theta$ scan [Fig.~\ref{Fig1}(c)], we observe all the allowed Bragg reflections associated with the (111) planes of LiV$_2$O$_4$, confirming that the film grew epitaxially with a well-defined orientation. We extract an interplane spacing of $d =$ 4.756 \AA, consistent with that of single crystals of LiV$_2$O$_4$, $d_{\textrm{bulk}} = a_{\textrm{bulk}}/\sqrt{3} = 8.240~\textrm{\AA}/\sqrt{3} \approx 4.758~\textrm{\AA}$ (Ref. \cite{Chmaissem_PRL_1997}). Reciprocal space mapping around the LiV$_2$O$_4$(226) peak [Fig.~\ref{Fig1}(d)] also reveals bulk-like values for the in-plane lattice parameter. Given the large lattice mismatch between bulk LiV$_2$O$_4$ and SrTiO$_3$, $(2a_{\textrm{SrTiO}_3} -  a_{\textrm{bulk}})/a_{\textrm{bulk}} = (7.810~\textrm{\AA}-a_{\textrm{bulk}})/a_{\textrm{bulk}} \approx -5.2\%$, we would expect a relaxed, dislocated growth of LiV$_2$O$_4$ on SrTiO$_3$. The relaxed growth may be responsible for the larger out-of-plane mosaicity of LiV$_2$O$_4$ films on SrTiO$_3$(111) compared to those on MgAl$_2$O$_4$(111) and MgO(001), which have lattice constants of $a_{\textrm{MgAl}_2\textrm{O}_4} = 8.085~\textrm{\AA}$ and $2a_{\textrm{MgO}} = 8.424~\textrm{\AA}$ and smaller lattice mismatches to LiV$_2$O$_4$, $-1.9\%$ and $2.2\%$, respectively. The larger out-of-plane mosaicity of the LiV$_2$O$_4$ films on SrTiO$_3$(111) is evidenced in the full-width half-maximum (FWHM) of the rocking curve, which is 0.71$^{\circ}$ as seen in the inset of Fig.~\ref{Fig1}(c), but 0.06$^{\circ}$ for films on both MgAl$_2$O$_4$(111) \cite{Yajima_PRB_2021} and MgO(001) \cite{Niemann_arXiv_2022}. Our films are metallic, as evidenced by transport measurements (Fig.~\ref{Fig1}(f); sample B). The resistivity exhibits the same qualitative behavior as in bulk single crystals~\cite{Urano_PRL_2000}: a characteristic downturn at $T^* \approx$ 20 K and a functional form $\rho = \rho_0 + AT^2$, $A$ = 4.5 $\mu \Omega$ cm/K$^2$, around 1.8 K (Fig.~\ref{Fig1}(f) inset). The residual resistivity ratio $\rho(\textrm{300 K}) / \rho(\textrm{0.5 K})$ of 3.5 is nearly identical to that of LiV$_2$O$_4$/MgAl$_2$O$_4$(111) films \cite{Yajima_PRB_2021}, in spite of the large mosaicity, but lower than the value of 11 in LiV$_2$O$_4$/SrTiO$_3$(001) films \cite{Niemann_arXiv_2022}. 

\begin{figure}
\includegraphics{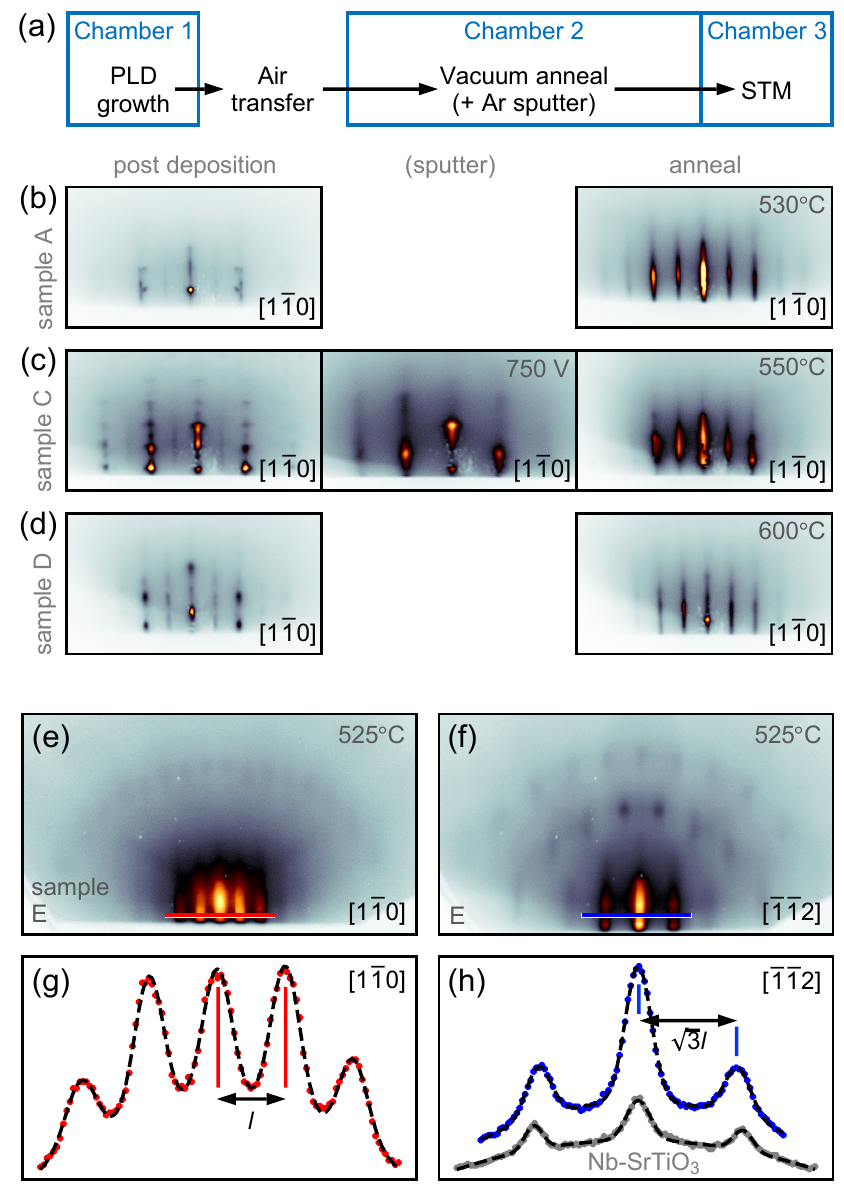}
\caption{\label{Fig2} (a) Diagram illustrating surface preparation of LiV$_2$O$_4$(111) films for STM measurements. (b)--(d) RHEED images of samples A, C, and D at various stages of preparation. Left: after PLD; middle (sample C only): Ar sputtering at 750 V; right: vacuum annealing at 530--600$^{\circ}$C. Electron energy: 15 keV. (e) and (f) Full RHEED images along the [1$\bar{1}$0] and [$\bar{1}\bar{1}$2] directions, acquired during vacuum annealing at 525$^{\circ}$C (sample E). Electron energy: 10 keV. Line cuts along the colored lines in (e) and (f) are shown in (g) and (h), respectively. The streak spacings of $l$ and $\sqrt{3}l$ indicate that the (111) surface has a hexagonal Bravais lattice. The dashed lines represent fits to a sum of Gaussian functions. A calibration curve acquired for Nb-SrTiO$_3$(111) along [$\bar{1}\bar{1}$2] is overlaid in (h).}
\end{figure}

\begin{figure*}
\includegraphics{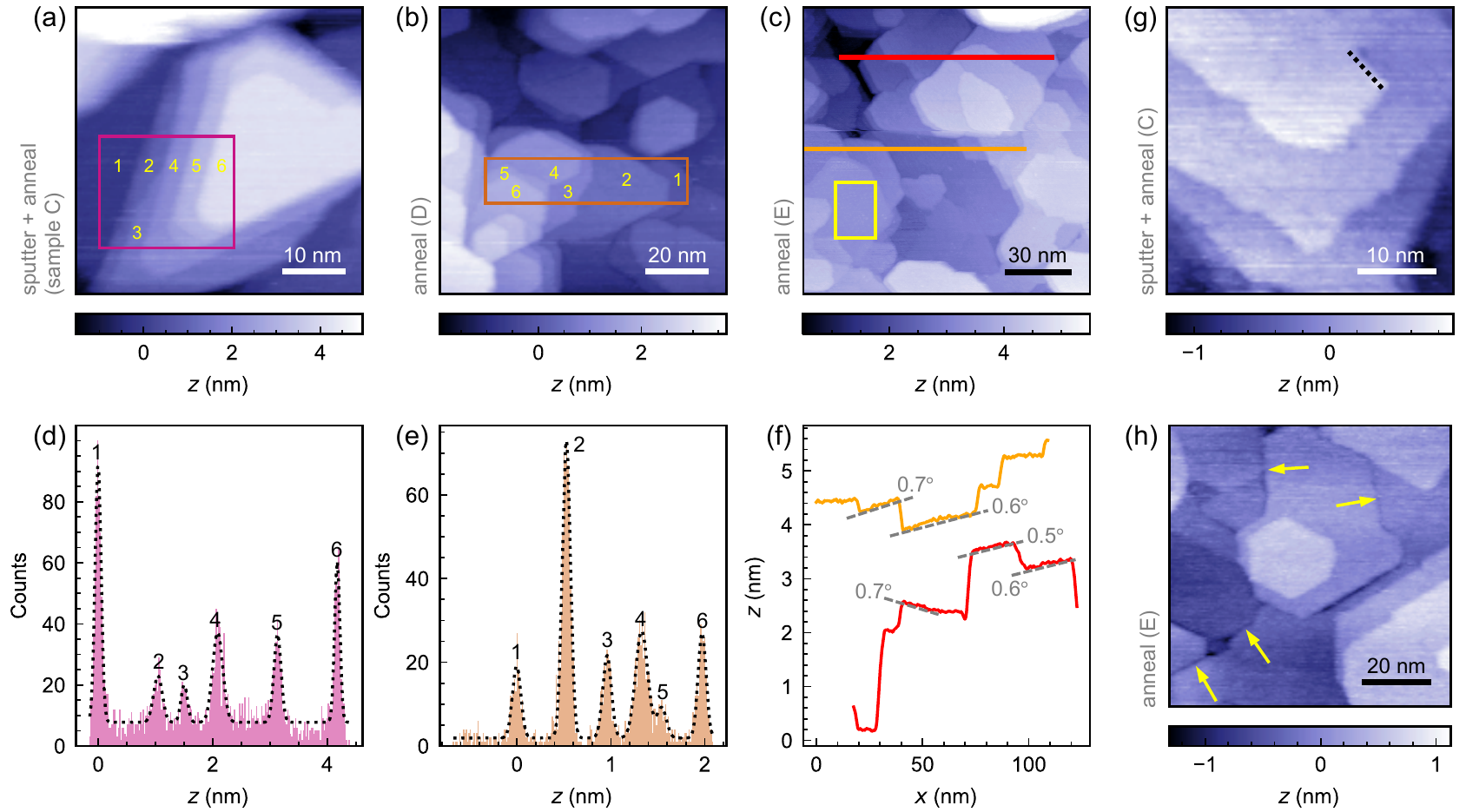}
\caption{\label{Fig3} (a)--(c) STM topographic images of LiV$_2$O$_4$. Setpoints: (a) 1 V, 20 pA, (b) 1 V, 30 pA, (c) 1 V, 50 pA. The root-mean-square roughness within the yellow box in (c) is 37 pm. (d) and (e) Histograms computed in the boxed areas of (a) and (b), respectively, with numbered peaks corresponding to the numbered terraces in (a) and (b). The dashed lines represent fits to a sum of Gaussian functions. In (d), the height difference between terraces 1 and 2, 2 and 4, 4 and 5, and 5 and 6 is 10.4~\textrm{\AA}, whereas the height difference between terraces 2 and 3 is 4.6~\textrm{\AA}. In (e), the height differences are 5.3~\textrm{\AA} between terraces 1 and 2, 4.3~\textrm{\AA} between terraces 2 and 3, 3.7~\textrm{\AA} between terraces 3 and 4, 2.1~\textrm{\AA} between terraces 4 and 5, and 4.3~\textrm{\AA} between terraces 5 and 6. (f) Height profiles along the red and orange lines in (c). The dashed lines mark terraces with an incline of 0.5--0.7$^{\circ}$. The orange line has been vertically offset for clarity. (g) and (h) Additional topographic images showing a screw dislocation (dotted line) and trenches between islands (arrows). Setpoints: (g) 1 V, 30 pA, (h) 1 V, 50 pA. The images in (a) and (g) were acquired for a film deposited on insulating SrTiO$_3$(111) (sample C, sputter and anneal), whereas the images in (b), (c), and (h) were acquired for films deposited on conducting Nb-SrTiO$_3$(111) (samples D and E, anneal only).}
\end{figure*}

Having confirmed that our thin films possess the desired LiV$_2$O$_4$ phase, we turn to STM measurements. The exposure of oxide thin films to air contaminates the surface and hampers atomically resolved STM. In particular, single crystals of LiV$_2$O$_4$ are known to be sensitive to air and moisture \cite{Matsushita_NatMat_2005}. Since we could not avoid atmospheric exposure when transferring our films from the PLD chamber to the STM system, we needed a means to recover pristine surfaces [Fig.~\ref{Fig2}(a)]. Furthermore, AFM reveals that as-grown films often show a granular topography [Fig.~\ref{Fig1}(e)]. To improve the surface quality, we performed Ar sputtering with moderate intensity (750 V, 10$^{-3}$ mbar Ar partial pressure), followed by annealing at 550$^{\circ}$C in UHV for a few hours (sample C). We could also obtain suitable surfaces with just UHV annealing at 530--600$^{\circ}$C, without sputtering (samples A, D, and E). Later STM images reveal no systematic differences between the two methods. As seen in Figs.~\ref{Fig2}(b)--\ref{Fig2}(d), the RHEED patterns of as-grown films are somewhat spotty, indicating a rougher, three-dimensional (3D) surface, but become brighter with sharper streaks after sputtering and annealing, indicating a smoother, 2D surface. The diffraction patterns are nearly identical to that of LiTi$_2$O$_4$(111) films \cite{Okada_NatComm_2017, Ohsawa_ACS_2020} and show clear in-plane orientation with streaks along [$\bar{1}$$\bar{1}$2] spaced $\sqrt{3}$ times farther apart than streaks along [1$\bar{1}$0], as expected for a hexagonal Bravais lattice [Figs.~\ref{Fig2}(e)--\ref{Fig2}(h)]. Comparing with the RHEED pattern obtained from a bare SrTiO$_3$(111) substrate [Fig.~\ref{Fig2}(h)], we extract the epitaxial relationship LiV$_2$O$_4$[$\bar{1}\bar{1}$2] $\parallel$ SrTiO$_3$[$\bar{1}\bar{1}$2] and a $-$4.9\% difference in the in-plane lattice constants, close to the $-$5.2\% mismatch of their bulk values. These results from surface-sensitive RHEED are fully consistent with those from bulk-sensitive XRD in Fig.~\ref{Fig1}. (Figures S4 and S5 in the supplementary material present XRD and transport measurements of a film after sputtering and annealing, confirming the target LiV$_2$O$_4$(111) phase with heavy fermion behavior, as well as an estimation of Li stoichiometry from the lattice constant.)

Figures \ref{Fig3}(a)--\ref{Fig3}(c) show room-temperature STM images of LiV$_2$O$_4$(111) films after annealing in UHV with or without sputtering, demonstrating clear step-and-terrace structure, in contrast to the granular morphology seen immediately after PLD [Fig.~\ref{Fig1}(e)]. We consistently observed a landscape of islands with atomically smooth tops. In Fig.~\ref{Fig3}(c), the root-mean-square roughness is 1.0 nm over a 140$\times$140 nm field of view with many islands, but 37 pm within a single island (yellow box). The islands have a typical lateral diameter of 20--30 nm and adopt compact shapes, including triangular [Fig.~\ref{Fig3}(a)], hexagonal [Figs.~\ref{Fig3}(b) and \ref{Fig3}(c)], and round [Figs.~\ref{Fig3}(b) and \ref{Fig3}(c)] shapes. These compact islands, as opposed to dendritic or fractal islands, indicate that the film adatoms have sufficient mobility to diffuse along the perimeter of the islands to achieve shapes that are more thermodynamically favored~\cite{Oura_2003}. Images of films with submonolayer coverage reveal that the formation of LiV$_2$O$_4$ islands begins at the SrTiO$_3$(111) interface (see Fig.~S6 in the supplementary material). The growth mode of LiV$_2$O$_4$ on SrTiO$_3$(111) is classified as Volmer-Weber, where the growth of islands starts at the substrate, as opposed to Stranski-Krastanov, where the film initially grows uniformly with layer-by-layer coverage, then switches to island growth beyond a critical thickness. Volmer-Weber growth mode is often facilitated by a large lattice mismatch between film and substrate, as is the case here.  

We take a closer look at the STM topographies by analyzing the island heights. In a few cases, such as in Fig.~\ref{Fig3}(a), where the islands on top of each other have the same shape with parallel step edges, we found the step heights to be regular and close to integer multiples of the lattice spacing along [111]. The histogram in Fig.~\ref{Fig3}(d) shows five terraces with a repeating height difference of 10.4~\textrm{\AA}, which is within 9\% of 2$d$, and a sixth terrace with a height difference 4.6~\textrm{\AA}, which is within 3\% of $d$. In many other cases, such as in Fig.~\ref{Fig3}(b), where the islands on top of each other have different shapes, the step heights are often irregular. The height differences between the successive terraces numbered in Fig.~\ref{Fig3}(b) are 5.3, 4.3, 3.7, 2.1, and 4.3~\textrm{\AA} [Fig.~\ref{Fig3}(e)]. These irregular step heights could imply that the islands have different surface terminations, as there are six distinct atomic layers along the [111] direction of LiV$_2$O$_4$ [Fig.~\ref{Fig1}(a)]. Alternatively, there could be stacking faults beneath the surface that shift the heights of the terraces by less than one full unit cell \cite{Goswami_SS_2007}. Spinel compounds are indeed known to host dislocations and antiphase boundaries that produce fractional shifts of the unit cell \cite{Yanina_SS_2002, Nedelkoski_SRep_2017}. In addition to the irregular terrace heights, we observed terraces that are tilted relative to the (111) plane. The height profiles in Fig.~\ref{Fig3}(f) reveal that some of the island tops are inclined, for example, by about 0.5--0.7$^{\circ}$. This out-of-plane mosaicity may correspond to the sizeable FWHM of the (111) rocking curve [inset of Fig.~\ref{Fig1}(c)].

We observed two further, distinct kinds of structural defects with STM: screw dislocations and trenches. Figure \ref{Fig3}(g) shows an example of an island with a screw dislocation (dotted line). The implication of a screw dislocation is that the island grows via spiral growth \cite{Goswami_SS_2007}. In spiral growth, adatoms nucleate at the exposed edge of the screw dislocation and cause the island to grow by winding. In contrast, the many other islands in Figs.~\ref{Fig3}(a)--\ref{Fig3}(c) with flat tops grow via 2D nucleation, wherein each layer is successively formed on top of another \cite{Redinger_PRL_2008}. From the prevalence of islands with flat tops over islands with screw dislocations in our images, we conclude that 2D nucleation plays a greater role than that of spiral growth in our films, but both growth modes are present. Figure \ref{Fig3}(h) shows examples of a different structural defect, that is, narrow trenches at the boundaries between islands. The implication of trenches is that the neighboring islands have failed to coalesce, likely because the edge atoms are out of phase. These boundaries make it difficult for the film to form one continuous domain and perpetuate island growth. 

\begin{figure}
\includegraphics{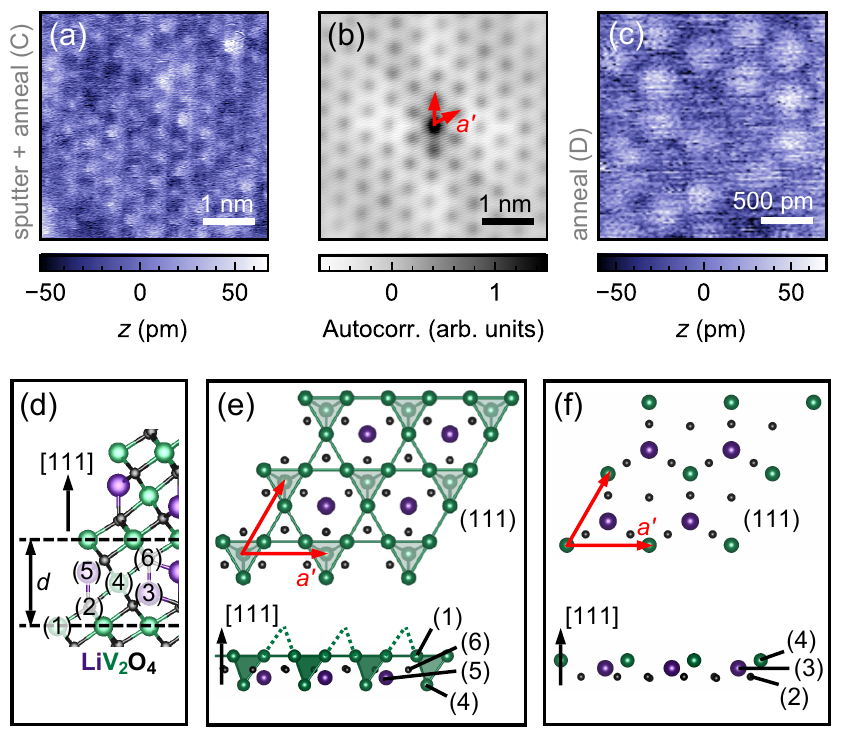}
\caption{\label{Fig4} (a) STM topographic image of LiV$_2$O$_4$(111) with atomic resolution (sample C, sputter and anneal). Setpoint: $-$76 mV, 45 pA. (b) Autocorrelation of (a). (c) Another topographic image (sample D, anneal only). Setpoint: 131 mV, 40 pA. (d) Six possible (111) terminations of LiV$_2$O$_4$, labeled (1)--(6). (e) Top and side views of the kagome (1)-V termination with several underlying layers. Complete V tetrahedra are shaded, whereas broken tetrahedra with missing V atoms are marked by the dotted line. (f) Top and side views of the hexagonal (4)-V termination with several underlying layers. In (e) and (f), the lattice constant $a'$ in the (111) plane is marked, which matches the periodicity observed in (b).}
\end{figure}

When the STM tip was brought closer to the surface of the islands, we observed a hexagonal atomic lattice [Figs.~\ref{Fig4}(a) and \ref{Fig4}(c)]. The lattice constant, as extracted from an autocorrelation [Fig.~\ref{Fig4}(b)], is $a' \approx 5.6$ \AA, close to and within the instrument $x$-$y$ calibration error of the bulk value for the LiV$_2$O$_4$(111)-1$\times$1 surface, $a_{\textrm{bulk}}/\sqrt{2} \approx 5.827$ \AA. Six atomic (111) terminations are possible, as labeled in Fig.~\ref{Fig4}(d): (1)-V, (2)-O, (3)-Li, (4)-V, (5)-Li, and (6)-O. We hypothesize that the bright lobes in the STM image arise from V atoms either in the kagome (1)-V [Fig.~\ref{Fig4}(e)] or hexagonal (4)-V [Fig.~\ref{Fig4}(f)] layers. First-principles calculations consistently find that the V $3d$ $t_{2g}$ orbitals dominate near the Fermi energy~\cite{Eyert_EPL_1999, Matsuno_PRB_1999}. Therefore, even if the surface were terminated by Li or O atoms, they could be invisible to STM. To further distinguish between the kagome (1)-V and hexagonal (4)-V terminations, however, is difficult. In the bulk spinel crystal, every V tetrahedron is equivalent. At a kagome (1)-V surface, this degeneracy is broken by a missing V atom in every second tetrahedron at the surface [dashed lines in Fig.~\ref{Fig4}(e)]. If the charge density becomes concentrated at the V tetrahedra that remain intact [shaded regions in Fig.~\ref{Fig4}(e)], then STM could also image a hexagonal lattice for the kagome (1)-V surface, similar to that of the hexagonal (4)-V surface.   

The stabilization of atomically flat LiV$_2$O$_4$(111) surfaces is a nontrivial observation whose mechanism requires further study. According to Tasker's classification~\cite{Tasker_JPC_1979}, the six possible atomic terminations in the (111) plane are all type III, meaning that a net dipole moment perpendicular to the surface exists. The resulting electrostatic potential would diverge, unless the surface atomic or electronic structure deviates from the bulk. Since we observe a LiV$_2$O$_4$(111)-1$\times$1 termination, we exclude a surface reconstruction that enlarges the unit cell in the in-plane directions, as often witnessed in the perovskite oxides, or [001]- and [110]-oriented spinels \cite{Chang_PRX_2016, Walls_PRB_2016}. Past studies of Fe$_3$O$_4$(111) and Co$_3$O$_4$(111) inferred that surface polarity could be compensated on a 1$\times$1 termination simply by an inward relaxation of the surface Fe$^{3+}$/Co$^{2+}$ ions and a corresponding shortening of the Fe-O/Co-O bonds, which indicates increased covalency~\cite{Ritter_SS_1999, Meyer_JPCM_2008}. Alternatively, Li deficiency~\cite{Azuma_JPCC_2013} (see Fig.~S5 in the supplementary material) or an exchange of the surface Li and V positions (inverse spinel structure), as proposed for LiMn$_2$O$_4$(111)~\cite{Kim_PRB_2015}, may also occur to counteract surface polarity. It is tempting to infer that these mechanisms to stabilize the polar (111) surface could also be connected to the irregular step heights observed in Fig. \ref{Fig3}. 

In summary, we have demonstrated that STM topographic imaging with atomic resolution is possible with epitaxial LiV$_2$O$_4$(111) films. An enabling step is a vacuum annealing procedure with optional Ar sputtering that cleanses and smoothens the surface. This simple technique circumvents the need for \textit{in-situ} transfer of films, thus opening the door to applying a wide range of surface-sensitive spectroscopic probes on LiV$_2$O$_4$. In particular, both low-temperature scanning tunneling spectroscopy and photoemission spectroscopy will be crucial in elucidating the nature of the heavy fermion phase. From our room-temperature STM measurements of LiV$_2$O$_4$, we could also glean clues about film-growth mechanisms. As a future work, it would be interesting to investigate whether uniform, layer-by-layer growth of LiV$_2$O$_4$ with fewer structural defects is possible by tuning growth and annealing conditions, choosing a substrate with a smaller lattice mismatch, or growing in the [001] or [110] orientations, or whether the island growth observed in this work is inevitable in the PLD of LiV$_2$O$_4$ films.   

\section*{Supplementary Material}

See supplementary material for additional information on growth and surface preparation, STM calibration, film properties after sputtering and annealing, estimation of Li stoichiometry, and Volmer-Weber growth mode of LiV$_2$O$_4$ on SrTiO$_3$(111).


%
%

%

\begin{acknowledgments}
We thank B. Stuhlhofer, G. Cristiani, K. Pflaum, M. Dueller, S. Prill-Diemer, H. Nakamura, K. K\"{u}ster, and U. Starke for technical support and use of facilities. D.H. acknowledges support from a Humboldt Research Fellowship for Postdoctoral Researchers.
\end{acknowledgments}


\begin{thebibliography}{49}%
\makeatletter
\providecommand \@ifxundefined [1]{%
 \@ifx{#1\undefined}
}%
\providecommand \@ifnum [1]{%
 \ifnum #1\expandafter \@firstoftwo
 \else \expandafter \@secondoftwo
 \fi
}%
\providecommand \@ifx [1]{%
 \ifx #1\expandafter \@firstoftwo
 \else \expandafter \@secondoftwo
 \fi
}%
\providecommand \natexlab [1]{#1}%
\providecommand \enquote  [1]{``#1''}%
\providecommand \bibnamefont  [1]{#1}%
\providecommand \bibfnamefont [1]{#1}%
\providecommand \citenamefont [1]{#1}%
\providecommand \href@noop [0]{\@secondoftwo}%
\providecommand \href [0]{\begingroup \@sanitize@url \@href}%
\providecommand \@href[1]{\@@startlink{#1}\@@href}%
\providecommand \@@href[1]{\endgroup#1\@@endlink}%
\providecommand \@sanitize@url [0]{\catcode `\\12\catcode `\$12\catcode
  `\&12\catcode `\#12\catcode `\^12\catcode `\_12\catcode `\%12\relax}%
\providecommand \@@startlink[1]{}%
\providecommand \@@endlink[0]{}%
\providecommand \url  [0]{\begingroup\@sanitize@url \@url }%
\providecommand \@url [1]{\endgroup\@href {#1}{\urlprefix }}%
\providecommand \urlprefix  [0]{URL }%
\providecommand \Eprint [0]{\href }%
\providecommand \doibase [0]{http://dx.doi.org/}%
\providecommand \selectlanguage [0]{\@gobble}%
\providecommand \bibinfo  [0]{\@secondoftwo}%
\providecommand \bibfield  [0]{\@secondoftwo}%
\providecommand \translation [1]{[#1]}%
\providecommand \BibitemOpen [0]{}%
\providecommand \bibitemStop [0]{}%
\providecommand \bibitemNoStop [0]{.\EOS\space}%
\providecommand \EOS [0]{\spacefactor3000\relax}%
\providecommand \BibitemShut  [1]{\csname bibitem#1\endcsname}%
\let\auto@bib@innerbib\@empty
\bibitem [{\citenamefont {Kondo}\ \emph {et~al.}(1997)\citenamefont {Kondo},
  \citenamefont {Johnston}, \citenamefont {Swenson}, \citenamefont {Borsa},
  \citenamefont {Mahajan}, \citenamefont {Miller}, \citenamefont {Gu},
  \citenamefont {Goldman}, \citenamefont {Maple}, \citenamefont {Gajewski},
  \citenamefont {Freeman}, \citenamefont {Dilley}, \citenamefont {Dickey},
  \citenamefont {Merrin}, \citenamefont {Kojima}, \citenamefont {Luke},
  \citenamefont {Uemura}, \citenamefont {Chmaissem},\ and\ \citenamefont
  {Jorgensen}}]{Kondo_PRL_1997}%
  \BibitemOpen
  \bibfield  {author} {\bibinfo {author} {\bibfnamefont {S.}~\bibnamefont
  {Kondo}}, \bibinfo {author} {\bibfnamefont {D.~C.}\ \bibnamefont {Johnston}},
  \bibinfo {author} {\bibfnamefont {C.~A.}\ \bibnamefont {Swenson}}, \bibinfo
  {author} {\bibfnamefont {F.}~\bibnamefont {Borsa}}, \bibinfo {author}
  {\bibfnamefont {A.~V.}\ \bibnamefont {Mahajan}}, \bibinfo {author}
  {\bibfnamefont {L.~L.}\ \bibnamefont {Miller}}, \bibinfo {author}
  {\bibfnamefont {T.}~\bibnamefont {Gu}}, \bibinfo {author} {\bibfnamefont
  {A.~I.}\ \bibnamefont {Goldman}}, \bibinfo {author} {\bibfnamefont {M.~B.}\
  \bibnamefont {Maple}}, \bibinfo {author} {\bibfnamefont {D.~A.}\ \bibnamefont
  {Gajewski}}, \bibinfo {author} {\bibfnamefont {E.~J.}\ \bibnamefont
  {Freeman}}, \bibinfo {author} {\bibfnamefont {N.~R.}\ \bibnamefont {Dilley}},
  \bibinfo {author} {\bibfnamefont {R.~P.}\ \bibnamefont {Dickey}}, \bibinfo
  {author} {\bibfnamefont {J.}~\bibnamefont {Merrin}}, \bibinfo {author}
  {\bibfnamefont {K.}~\bibnamefont {Kojima}}, \bibinfo {author} {\bibfnamefont
  {G.~M.}\ \bibnamefont {Luke}}, \bibinfo {author} {\bibfnamefont {Y.~J.}\
  \bibnamefont {Uemura}}, \bibinfo {author} {\bibfnamefont {O.}~\bibnamefont
  {Chmaissem}}, \ and\ \bibinfo {author} {\bibfnamefont {J.~D.}\ \bibnamefont
  {Jorgensen}},\ }\href {\doibase 10.1103/PhysRevLett.78.3729} {\bibfield
  {journal} {\bibinfo  {journal} {Phys. Rev. Lett.}\ }\textbf {\bibinfo
  {volume} {78}},\ \bibinfo {pages} {3729} (\bibinfo {year}
  {1997})}\BibitemShut {NoStop}%
\bibitem [{\citenamefont {Urano}\ \emph {et~al.}(2000)\citenamefont {Urano},
  \citenamefont {Nohara}, \citenamefont {Kondo}, \citenamefont {Sakai},
  \citenamefont {Takagi}, \citenamefont {Shiraki},\ and\ \citenamefont
  {Okubo}}]{Urano_PRL_2000}%
  \BibitemOpen
  \bibfield  {author} {\bibinfo {author} {\bibfnamefont {C.}~\bibnamefont
  {Urano}}, \bibinfo {author} {\bibfnamefont {M.}~\bibnamefont {Nohara}},
  \bibinfo {author} {\bibfnamefont {S.}~\bibnamefont {Kondo}}, \bibinfo
  {author} {\bibfnamefont {F.}~\bibnamefont {Sakai}}, \bibinfo {author}
  {\bibfnamefont {H.}~\bibnamefont {Takagi}}, \bibinfo {author} {\bibfnamefont
  {T.}~\bibnamefont {Shiraki}}, \ and\ \bibinfo {author} {\bibfnamefont
  {T.}~\bibnamefont {Okubo}},\ }\href {\doibase 10.1103/PhysRevLett.85.1052}
  {\bibfield  {journal} {\bibinfo  {journal} {Phys. Rev. Lett.}\ }\textbf
  {\bibinfo {volume} {85}},\ \bibinfo {pages} {1052} (\bibinfo {year}
  {2000})}\BibitemShut {NoStop}%
\bibitem [{\citenamefont {Kessler}\ and\ \citenamefont
  {Sienko}(1971)}]{Kessler_JCP_1971}%
  \BibitemOpen
  \bibfield  {author} {\bibinfo {author} {\bibfnamefont {H.}~\bibnamefont
  {Kessler}}\ and\ \bibinfo {author} {\bibfnamefont {M.~J.}\ \bibnamefont
  {Sienko}},\ }\href {\doibase 10.1063/1.1675694} {\bibfield  {journal}
  {\bibinfo  {journal} {J. Chem. Phys.}\ }\textbf {\bibinfo {volume} {55}},\
  \bibinfo {pages} {5414} (\bibinfo {year} {1971})}\BibitemShut {NoStop}%
\bibitem [{\citenamefont {Anderson}(1956)}]{Anderson_PR_1956}%
  \BibitemOpen
  \bibfield  {author} {\bibinfo {author} {\bibfnamefont {P.~W.}\ \bibnamefont
  {Anderson}},\ }\href {\doibase 10.1103/PhysRev.102.1008} {\bibfield
  {journal} {\bibinfo  {journal} {Phys. Rev.}\ }\textbf {\bibinfo {volume}
  {102}},\ \bibinfo {pages} {1008} (\bibinfo {year} {1956})}\BibitemShut
  {NoStop}%
\bibitem [{\citenamefont {Verwey}(1939)}]{Verwey_Nature_1939}%
  \BibitemOpen
  \bibfield  {author} {\bibinfo {author} {\bibfnamefont {E.~J.~W.}\
  \bibnamefont {Verwey}},\ }\href {\doibase 10.1038/144327b0} {\bibfield
  {journal} {\bibinfo  {journal} {Nature}\ }\textbf {\bibinfo {volume} {144}},\
  \bibinfo {pages} {327} (\bibinfo {year} {1939})}\BibitemShut {NoStop}%
\bibitem [{\citenamefont {Matsuno}\ \emph {et~al.}(2001)\citenamefont
  {Matsuno}, \citenamefont {Katsufuji}, \citenamefont {Mori}, \citenamefont
  {Moritomo}, \citenamefont {Machida}, \citenamefont {Nishibori}, \citenamefont
  {Takata}, \citenamefont {Sakata}, \citenamefont {Yamamoto},\ and\
  \citenamefont {Takagi}}]{Matsuno_JPSJ_2001}%
  \BibitemOpen
  \bibfield  {author} {\bibinfo {author} {\bibfnamefont {K.}~\bibnamefont
  {Matsuno}}, \bibinfo {author} {\bibfnamefont {T.}~\bibnamefont {Katsufuji}},
  \bibinfo {author} {\bibfnamefont {S.}~\bibnamefont {Mori}}, \bibinfo {author}
  {\bibfnamefont {Y.}~\bibnamefont {Moritomo}}, \bibinfo {author}
  {\bibfnamefont {A.}~\bibnamefont {Machida}}, \bibinfo {author} {\bibfnamefont
  {E.}~\bibnamefont {Nishibori}}, \bibinfo {author} {\bibfnamefont
  {M.}~\bibnamefont {Takata}}, \bibinfo {author} {\bibfnamefont
  {M.}~\bibnamefont {Sakata}}, \bibinfo {author} {\bibfnamefont
  {N.}~\bibnamefont {Yamamoto}}, \ and\ \bibinfo {author} {\bibfnamefont
  {H.}~\bibnamefont {Takagi}},\ }\href {\doibase 10.1143/JPSJ.70.1456}
  {\bibfield  {journal} {\bibinfo  {journal} {J. Phys. Soc. Japan}\ }\textbf
  {\bibinfo {volume} {70}},\ \bibinfo {pages} {1456} (\bibinfo {year}
  {2001})}\BibitemShut {NoStop}%
\bibitem [{\citenamefont {Yamada}\ and\ \citenamefont
  {Tanaka}(1995)}]{Yamada_MRB_1995}%
  \BibitemOpen
  \bibfield  {author} {\bibinfo {author} {\bibfnamefont {A.}~\bibnamefont
  {Yamada}}\ and\ \bibinfo {author} {\bibfnamefont {M.}~\bibnamefont
  {Tanaka}},\ }\href {\doibase https://doi.org/10.1016/0025-5408(95)00048-8}
  {\bibfield  {journal} {\bibinfo  {journal} {Mater. Res. Bull.}\ }\textbf
  {\bibinfo {volume} {30}},\ \bibinfo {pages} {715} (\bibinfo {year}
  {1995})}\BibitemShut {NoStop}%
\bibitem [{\citenamefont {Okamoto}\ \emph {et~al.}(2008)\citenamefont
  {Okamoto}, \citenamefont {Niitaka}, \citenamefont {Uchida}, \citenamefont
  {Waki}, \citenamefont {Takigawa}, \citenamefont {Nakatsu}, \citenamefont
  {Sekiyama}, \citenamefont {Suga}, \citenamefont {Arita},\ and\ \citenamefont
  {Takagi}}]{Okamoto_PRL_2008}%
  \BibitemOpen
  \bibfield  {author} {\bibinfo {author} {\bibfnamefont {Y.}~\bibnamefont
  {Okamoto}}, \bibinfo {author} {\bibfnamefont {S.}~\bibnamefont {Niitaka}},
  \bibinfo {author} {\bibfnamefont {M.}~\bibnamefont {Uchida}}, \bibinfo
  {author} {\bibfnamefont {T.}~\bibnamefont {Waki}}, \bibinfo {author}
  {\bibfnamefont {M.}~\bibnamefont {Takigawa}}, \bibinfo {author}
  {\bibfnamefont {Y.}~\bibnamefont {Nakatsu}}, \bibinfo {author} {\bibfnamefont
  {A.}~\bibnamefont {Sekiyama}}, \bibinfo {author} {\bibfnamefont
  {S.}~\bibnamefont {Suga}}, \bibinfo {author} {\bibfnamefont {R.}~\bibnamefont
  {Arita}}, \ and\ \bibinfo {author} {\bibfnamefont {H.}~\bibnamefont
  {Takagi}},\ }\href {\doibase 10.1103/PhysRevLett.101.086404} {\bibfield
  {journal} {\bibinfo  {journal} {Phys. Rev. Lett.}\ }\textbf {\bibinfo
  {volume} {101}},\ \bibinfo {pages} {086404} (\bibinfo {year}
  {2008})}\BibitemShut {NoStop}%
\bibitem [{\citenamefont {Browne}\ \emph {et~al.}(2018)\citenamefont {Browne},
  \citenamefont {Lithgow}, \citenamefont {Kimber},\ and\ \citenamefont
  {Attfield}}]{Browne_IC_2018}%
  \BibitemOpen
  \bibfield  {author} {\bibinfo {author} {\bibfnamefont {A.~J.}\ \bibnamefont
  {Browne}}, \bibinfo {author} {\bibfnamefont {C.}~\bibnamefont {Lithgow}},
  \bibinfo {author} {\bibfnamefont {S.~A.~J.}\ \bibnamefont {Kimber}}, \ and\
  \bibinfo {author} {\bibfnamefont {J.~P.}\ \bibnamefont {Attfield}},\ }\href
  {\doibase 10.1021/acs.inorgchem.7b03221} {\bibfield  {journal} {\bibinfo
  {journal} {Inorg. Chem.}\ }\textbf {\bibinfo {volume} {57}},\ \bibinfo
  {pages} {2815} (\bibinfo {year} {2018})}\BibitemShut {NoStop}%
\bibitem [{\citenamefont {Chmaissem}\ \emph {et~al.}(1997)\citenamefont
  {Chmaissem}, \citenamefont {Jorgensen}, \citenamefont {Kondo},\ and\
  \citenamefont {Johnston}}]{Chmaissem_PRL_1997}%
  \BibitemOpen
  \bibfield  {author} {\bibinfo {author} {\bibfnamefont {O.}~\bibnamefont
  {Chmaissem}}, \bibinfo {author} {\bibfnamefont {J.~D.}\ \bibnamefont
  {Jorgensen}}, \bibinfo {author} {\bibfnamefont {S.}~\bibnamefont {Kondo}}, \
  and\ \bibinfo {author} {\bibfnamefont {D.~C.}\ \bibnamefont {Johnston}},\
  }\href {\doibase 10.1103/PhysRevLett.79.4866} {\bibfield  {journal} {\bibinfo
   {journal} {Phys. Rev. Lett.}\ }\textbf {\bibinfo {volume} {79}},\ \bibinfo
  {pages} {4866} (\bibinfo {year} {1997})}\BibitemShut {NoStop}%
\bibitem [{\citenamefont {Matsuno}, \citenamefont {Fujimori},\ and\
  \citenamefont {Mattheiss}(1999)}]{Matsuno_PRB_1999}%
  \BibitemOpen
  \bibfield  {author} {\bibinfo {author} {\bibfnamefont {J.}~\bibnamefont
  {Matsuno}}, \bibinfo {author} {\bibfnamefont {A.}~\bibnamefont {Fujimori}}, \
  and\ \bibinfo {author} {\bibfnamefont {L.~F.}\ \bibnamefont {Mattheiss}},\
  }\href {\doibase 10.1103/PhysRevB.60.1607} {\bibfield  {journal} {\bibinfo
  {journal} {Phys. Rev. B}\ }\textbf {\bibinfo {volume} {60}},\ \bibinfo
  {pages} {1607} (\bibinfo {year} {1999})}\BibitemShut {NoStop}%
\bibitem [{\citenamefont {Shimoyamada}\ \emph {et~al.}(2006)\citenamefont
  {Shimoyamada}, \citenamefont {Tsuda}, \citenamefont {Ishizaka}, \citenamefont
  {Kiss}, \citenamefont {Shimojima}, \citenamefont {Togashi}, \citenamefont
  {Watanabe}, \citenamefont {Zhang}, \citenamefont {Chen}, \citenamefont
  {Matsushita}, \citenamefont {Ueda}, \citenamefont {Ueda},\ and\ \citenamefont
  {Shin}}]{Shimoyamada_PRL_2006}%
  \BibitemOpen
  \bibfield  {author} {\bibinfo {author} {\bibfnamefont {A.}~\bibnamefont
  {Shimoyamada}}, \bibinfo {author} {\bibfnamefont {S.}~\bibnamefont {Tsuda}},
  \bibinfo {author} {\bibfnamefont {K.}~\bibnamefont {Ishizaka}}, \bibinfo
  {author} {\bibfnamefont {T.}~\bibnamefont {Kiss}}, \bibinfo {author}
  {\bibfnamefont {T.}~\bibnamefont {Shimojima}}, \bibinfo {author}
  {\bibfnamefont {T.}~\bibnamefont {Togashi}}, \bibinfo {author} {\bibfnamefont
  {S.}~\bibnamefont {Watanabe}}, \bibinfo {author} {\bibfnamefont {C.~Q.}\
  \bibnamefont {Zhang}}, \bibinfo {author} {\bibfnamefont {C.~T.}\ \bibnamefont
  {Chen}}, \bibinfo {author} {\bibfnamefont {Y.}~\bibnamefont {Matsushita}},
  \bibinfo {author} {\bibfnamefont {H.}~\bibnamefont {Ueda}}, \bibinfo {author}
  {\bibfnamefont {Y.}~\bibnamefont {Ueda}}, \ and\ \bibinfo {author}
  {\bibfnamefont {S.}~\bibnamefont {Shin}},\ }\href {\doibase
  10.1103/PhysRevLett.96.026403} {\bibfield  {journal} {\bibinfo  {journal}
  {Phys. Rev. Lett.}\ }\textbf {\bibinfo {volume} {96}},\ \bibinfo {pages}
  {026403} (\bibinfo {year} {2006})}\BibitemShut {NoStop}%
\bibitem [{\citenamefont {J\"onsson}\ \emph {et~al.}(2007)\citenamefont
  {J\"onsson}, \citenamefont {Takenaka}, \citenamefont {Niitaka}, \citenamefont
  {Sasagawa}, \citenamefont {Sugai},\ and\ \citenamefont
  {Takagi}}]{Jonsson_PRL_2007}%
  \BibitemOpen
  \bibfield  {author} {\bibinfo {author} {\bibfnamefont {P.~E.}\ \bibnamefont
  {J\"onsson}}, \bibinfo {author} {\bibfnamefont {K.}~\bibnamefont {Takenaka}},
  \bibinfo {author} {\bibfnamefont {S.}~\bibnamefont {Niitaka}}, \bibinfo
  {author} {\bibfnamefont {T.}~\bibnamefont {Sasagawa}}, \bibinfo {author}
  {\bibfnamefont {S.}~\bibnamefont {Sugai}}, \ and\ \bibinfo {author}
  {\bibfnamefont {H.}~\bibnamefont {Takagi}},\ }\href {\doibase
  10.1103/PhysRevLett.99.167402} {\bibfield  {journal} {\bibinfo  {journal}
  {Phys. Rev. Lett.}\ }\textbf {\bibinfo {volume} {99}},\ \bibinfo {pages}
  {167402} (\bibinfo {year} {2007})}\BibitemShut {NoStop}%
\bibitem [{\citenamefont {Shimizu}\ \emph {et~al.}(2012)\citenamefont
  {Shimizu}, \citenamefont {Takeda}, \citenamefont {Tanaka}, \citenamefont
  {Itoh}, \citenamefont {Niitaka},\ and\ \citenamefont
  {Takagi}}]{Shimizu_NatComm_2012}%
  \BibitemOpen
  \bibfield  {author} {\bibinfo {author} {\bibfnamefont {Y.}~\bibnamefont
  {Shimizu}}, \bibinfo {author} {\bibfnamefont {H.}~\bibnamefont {Takeda}},
  \bibinfo {author} {\bibfnamefont {M.}~\bibnamefont {Tanaka}}, \bibinfo
  {author} {\bibfnamefont {M.}~\bibnamefont {Itoh}}, \bibinfo {author}
  {\bibfnamefont {S.}~\bibnamefont {Niitaka}}, \ and\ \bibinfo {author}
  {\bibfnamefont {H.}~\bibnamefont {Takagi}},\ }\href {\doibase
  10.1038/ncomms1979} {\bibfield  {journal} {\bibinfo  {journal} {Nat.
  Commun.}\ }\textbf {\bibinfo {volume} {3}},\ \bibinfo {pages} {981} (\bibinfo
  {year} {2012})}\BibitemShut {NoStop}%
\bibitem [{\citenamefont {Browne}\ \emph {et~al.}(2020)\citenamefont {Browne},
  \citenamefont {Pace}, \citenamefont {Garbarino},\ and\ \citenamefont
  {Attfield}}]{Browne_PRM_2020}%
  \BibitemOpen
  \bibfield  {author} {\bibinfo {author} {\bibfnamefont {A.~J.}\ \bibnamefont
  {Browne}}, \bibinfo {author} {\bibfnamefont {E.~J.}\ \bibnamefont {Pace}},
  \bibinfo {author} {\bibfnamefont {G.}~\bibnamefont {Garbarino}}, \ and\
  \bibinfo {author} {\bibfnamefont {J.~P.}\ \bibnamefont {Attfield}},\ }\href
  {\doibase 10.1103/PhysRevMaterials.4.015002} {\bibfield  {journal} {\bibinfo
  {journal} {Phys. Rev. Mater.}\ }\textbf {\bibinfo {volume} {4}},\ \bibinfo
  {pages} {015002} (\bibinfo {year} {2020})}\BibitemShut {NoStop}%
\bibitem [{\citenamefont {Anisimov}\ \emph {et~al.}(1999)\citenamefont
  {Anisimov}, \citenamefont {Korotin}, \citenamefont {Z\"olfl}, \citenamefont
  {Pruschke}, \citenamefont {Le~Hur},\ and\ \citenamefont
  {Rice}}]{Anisimov_PRL_1999}%
  \BibitemOpen
  \bibfield  {author} {\bibinfo {author} {\bibfnamefont {V.~I.}\ \bibnamefont
  {Anisimov}}, \bibinfo {author} {\bibfnamefont {M.~A.}\ \bibnamefont
  {Korotin}}, \bibinfo {author} {\bibfnamefont {M.}~\bibnamefont {Z\"olfl}},
  \bibinfo {author} {\bibfnamefont {T.}~\bibnamefont {Pruschke}}, \bibinfo
  {author} {\bibfnamefont {K.}~\bibnamefont {Le~Hur}}, \ and\ \bibinfo {author}
  {\bibfnamefont {T.~M.}\ \bibnamefont {Rice}},\ }\href {\doibase
  10.1103/PhysRevLett.83.364} {\bibfield  {journal} {\bibinfo  {journal} {Phys.
  Rev. Lett.}\ }\textbf {\bibinfo {volume} {83}},\ \bibinfo {pages} {364}
  (\bibinfo {year} {1999})}\BibitemShut {NoStop}%
\bibitem [{\citenamefont {Fulde}\ \emph {et~al.}(2001)\citenamefont {Fulde},
  \citenamefont {Yaresko}, \citenamefont {Zvyagin},\ and\ \citenamefont
  {Grin}}]{Fulde_EPL_2001}%
  \BibitemOpen
  \bibfield  {author} {\bibinfo {author} {\bibfnamefont {P.}~\bibnamefont
  {Fulde}}, \bibinfo {author} {\bibfnamefont {A.~N.}\ \bibnamefont {Yaresko}},
  \bibinfo {author} {\bibfnamefont {A.~A.}\ \bibnamefont {Zvyagin}}, \ and\
  \bibinfo {author} {\bibfnamefont {Y.}~\bibnamefont {Grin}},\ }\href {\doibase
  10.1209/epl/i2001-00322-3} {\bibfield  {journal} {\bibinfo  {journal} {EPL}\
  }\textbf {\bibinfo {volume} {54}},\ \bibinfo {pages} {779} (\bibinfo {year}
  {2001})}\BibitemShut {NoStop}%
\bibitem [{\citenamefont {Burdin}, \citenamefont {Grempel},\ and\ \citenamefont
  {Georges}(2002)}]{Burdin_PRB_2002}%
  \BibitemOpen
  \bibfield  {author} {\bibinfo {author} {\bibfnamefont {S.}~\bibnamefont
  {Burdin}}, \bibinfo {author} {\bibfnamefont {D.~R.}\ \bibnamefont {Grempel}},
  \ and\ \bibinfo {author} {\bibfnamefont {A.}~\bibnamefont {Georges}},\ }\href
  {\doibase 10.1103/PhysRevB.66.045111} {\bibfield  {journal} {\bibinfo
  {journal} {Phys. Rev. B}\ }\textbf {\bibinfo {volume} {66}},\ \bibinfo
  {pages} {045111} (\bibinfo {year} {2002})}\BibitemShut {NoStop}%
\bibitem [{\citenamefont {Fujimoto}(2002)}]{Fujimoto_PRB_2002}%
  \BibitemOpen
  \bibfield  {author} {\bibinfo {author} {\bibfnamefont {S.}~\bibnamefont
  {Fujimoto}},\ }\href {\doibase 10.1103/PhysRevB.65.155108} {\bibfield
  {journal} {\bibinfo  {journal} {Phys. Rev. B}\ }\textbf {\bibinfo {volume}
  {65}},\ \bibinfo {pages} {155108} (\bibinfo {year} {2002})}\BibitemShut
  {NoStop}%
\bibitem [{\citenamefont {Hopkinson}\ and\ \citenamefont
  {Coleman}(2002)}]{Hopkinson_PRL_2002}%
  \BibitemOpen
  \bibfield  {author} {\bibinfo {author} {\bibfnamefont {J.}~\bibnamefont
  {Hopkinson}}\ and\ \bibinfo {author} {\bibfnamefont {P.}~\bibnamefont
  {Coleman}},\ }\href {\doibase 10.1103/PhysRevLett.89.267201} {\bibfield
  {journal} {\bibinfo  {journal} {Phys. Rev. Lett.}\ }\textbf {\bibinfo
  {volume} {89}},\ \bibinfo {pages} {267201} (\bibinfo {year}
  {2002})}\BibitemShut {NoStop}%
\bibitem [{\citenamefont {Yamashita}\ and\ \citenamefont
  {Ueda}(2003)}]{Yamashita_PRB_2003}%
  \BibitemOpen
  \bibfield  {author} {\bibinfo {author} {\bibfnamefont {Y.}~\bibnamefont
  {Yamashita}}\ and\ \bibinfo {author} {\bibfnamefont {K.}~\bibnamefont
  {Ueda}},\ }\href {\doibase 10.1103/PhysRevB.67.195107} {\bibfield  {journal}
  {\bibinfo  {journal} {Phys. Rev. B}\ }\textbf {\bibinfo {volume} {67}},\
  \bibinfo {pages} {195107} (\bibinfo {year} {2003})}\BibitemShut {NoStop}%
\bibitem [{\citenamefont {Arita}\ \emph {et~al.}(2007)\citenamefont {Arita},
  \citenamefont {Held}, \citenamefont {Lukoyanov},\ and\ \citenamefont
  {Anisimov}}]{Arita_PRL_2007}%
  \BibitemOpen
  \bibfield  {author} {\bibinfo {author} {\bibfnamefont {R.}~\bibnamefont
  {Arita}}, \bibinfo {author} {\bibfnamefont {K.}~\bibnamefont {Held}},
  \bibinfo {author} {\bibfnamefont {A.~V.}\ \bibnamefont {Lukoyanov}}, \ and\
  \bibinfo {author} {\bibfnamefont {V.~I.}\ \bibnamefont {Anisimov}},\ }\href
  {\doibase 10.1103/PhysRevLett.98.166402} {\bibfield  {journal} {\bibinfo
  {journal} {Phys. Rev. Lett.}\ }\textbf {\bibinfo {volume} {98}},\ \bibinfo
  {pages} {166402} (\bibinfo {year} {2007})}\BibitemShut {NoStop}%
\bibitem [{\citenamefont {Hattori}\ and\ \citenamefont
  {Tsunetsugu}(2009)}]{Hattori_PRB_2009}%
  \BibitemOpen
  \bibfield  {author} {\bibinfo {author} {\bibfnamefont {K.}~\bibnamefont
  {Hattori}}\ and\ \bibinfo {author} {\bibfnamefont {H.}~\bibnamefont
  {Tsunetsugu}},\ }\href {\doibase 10.1103/PhysRevB.79.035115} {\bibfield
  {journal} {\bibinfo  {journal} {Phys. Rev. B}\ }\textbf {\bibinfo {volume}
  {79}},\ \bibinfo {pages} {035115} (\bibinfo {year} {2009})}\BibitemShut
  {NoStop}%
\bibitem [{\citenamefont {Yajima}\ \emph {et~al.}(2021)\citenamefont {Yajima},
  \citenamefont {Soma}, \citenamefont {Yoshimatsu}, \citenamefont {Kurita},
  \citenamefont {Watanabe},\ and\ \citenamefont {Ohtomo}}]{Yajima_PRB_2021}%
  \BibitemOpen
  \bibfield  {author} {\bibinfo {author} {\bibfnamefont {T.}~\bibnamefont
  {Yajima}}, \bibinfo {author} {\bibfnamefont {T.}~\bibnamefont {Soma}},
  \bibinfo {author} {\bibfnamefont {K.}~\bibnamefont {Yoshimatsu}}, \bibinfo
  {author} {\bibfnamefont {N.}~\bibnamefont {Kurita}}, \bibinfo {author}
  {\bibfnamefont {M.}~\bibnamefont {Watanabe}}, \ and\ \bibinfo {author}
  {\bibfnamefont {A.}~\bibnamefont {Ohtomo}},\ }\href {\doibase
  10.1103/PhysRevB.104.245104} {\bibfield  {journal} {\bibinfo  {journal}
  {Phys. Rev. B}\ }\textbf {\bibinfo {volume} {104}},\ \bibinfo {pages}
  {245104} (\bibinfo {year} {2021})}\BibitemShut {NoStop}%
\bibitem [{\citenamefont {Niemann}\ \emph {et~al.}()\citenamefont {Niemann},
  \citenamefont {Wu}, \citenamefont {Oka}, \citenamefont {Hirai}, \citenamefont
  {Wang}, \citenamefont {Suyolcu}, \citenamefont {Kim}, \citenamefont {van
  Aken},\ and\ \citenamefont {Takagi}}]{Niemann_arXiv_2022}%
  \BibitemOpen
  \bibfield  {author} {\bibinfo {author} {\bibfnamefont {U.}~\bibnamefont
  {Niemann}}, \bibinfo {author} {\bibfnamefont {Y.~M.}\ \bibnamefont {Wu}},
  \bibinfo {author} {\bibfnamefont {R.}~\bibnamefont {Oka}}, \bibinfo {author}
  {\bibfnamefont {D.}~\bibnamefont {Hirai}}, \bibinfo {author} {\bibfnamefont
  {Y.}~\bibnamefont {Wang}}, \bibinfo {author} {\bibfnamefont {Y.~E.}\
  \bibnamefont {Suyolcu}}, \bibinfo {author} {\bibfnamefont {M.}~\bibnamefont
  {Kim}}, \bibinfo {author} {\bibfnamefont {P.~A.}\ \bibnamefont {van Aken}}, \
  and\ \bibinfo {author} {\bibfnamefont {H.}~\bibnamefont {Takagi}},\ }\href
  {\doibase 10.48550/ARXIV.2206.11585} {}\Eprint
  {http://arxiv.org/abs/2206.11585} {arXiv:2206.11585} \BibitemShut {NoStop}%
\bibitem [{\citenamefont {Lu}\ \emph {et~al.}(2019)\citenamefont {Lu},
  \citenamefont {Zeng}, \citenamefont {Wang}, \citenamefont {Yang},
  \citenamefont {Hu}, \citenamefont {Jia}, \citenamefont {Zhao}, \citenamefont
  {Yin}, \citenamefont {Ge},\ and\ \citenamefont {Xi}}]{Lu_AMI_2019}%
  \BibitemOpen
  \bibfield  {author} {\bibinfo {author} {\bibfnamefont {Y.}~\bibnamefont
  {Lu}}, \bibinfo {author} {\bibfnamefont {X.}~\bibnamefont {Zeng}}, \bibinfo
  {author} {\bibfnamefont {J.}~\bibnamefont {Wang}}, \bibinfo {author}
  {\bibfnamefont {L.}~\bibnamefont {Yang}}, \bibinfo {author} {\bibfnamefont
  {S.}~\bibnamefont {Hu}}, \bibinfo {author} {\bibfnamefont {C.}~\bibnamefont
  {Jia}}, \bibinfo {author} {\bibfnamefont {H.}~\bibnamefont {Zhao}}, \bibinfo
  {author} {\bibfnamefont {D.}~\bibnamefont {Yin}}, \bibinfo {author}
  {\bibfnamefont {X.}~\bibnamefont {Ge}}, \ and\ \bibinfo {author}
  {\bibfnamefont {X.}~\bibnamefont {Xi}},\ }\href {\doibase
  https://doi.org/10.1002/admi.201901368} {\bibfield  {journal} {\bibinfo
  {journal} {Adv. Mater. Interfaces}\ }\textbf {\bibinfo {volume} {6}},\
  \bibinfo {pages} {1901368} (\bibinfo {year} {2019})}\BibitemShut {NoStop}%
\bibitem [{\citenamefont {Das}\ \emph {et~al.}(2007)\citenamefont {Das},
  \citenamefont {Zong}, \citenamefont {Niazi}, \citenamefont {Ellern},
  \citenamefont {Yan},\ and\ \citenamefont {Johnston}}]{Das_PRB_2007}%
  \BibitemOpen
  \bibfield  {author} {\bibinfo {author} {\bibfnamefont {S.}~\bibnamefont
  {Das}}, \bibinfo {author} {\bibfnamefont {X.}~\bibnamefont {Zong}}, \bibinfo
  {author} {\bibfnamefont {A.}~\bibnamefont {Niazi}}, \bibinfo {author}
  {\bibfnamefont {A.}~\bibnamefont {Ellern}}, \bibinfo {author} {\bibfnamefont
  {J.~Q.}\ \bibnamefont {Yan}}, \ and\ \bibinfo {author} {\bibfnamefont
  {D.~C.}\ \bibnamefont {Johnston}},\ }\href {\doibase
  10.1103/PhysRevB.76.054418} {\bibfield  {journal} {\bibinfo  {journal} {Phys.
  Rev. B}\ }\textbf {\bibinfo {volume} {76}},\ \bibinfo {pages} {054418}
  (\bibinfo {year} {2007})}\BibitemShut {NoStop}%
\bibitem [{\citenamefont {Ritter}\ and\ \citenamefont
  {Weiss}(1999)}]{Ritter_SS_1999}%
  \BibitemOpen
  \bibfield  {author} {\bibinfo {author} {\bibfnamefont {M.}~\bibnamefont
  {Ritter}}\ and\ \bibinfo {author} {\bibfnamefont {W.}~\bibnamefont {Weiss}},\
  }\href {\doibase https://doi.org/10.1016/S0039-6028(99)00518-X} {\bibfield
  {journal} {\bibinfo  {journal} {Surf. Sci.}\ }\textbf {\bibinfo {volume}
  {432}},\ \bibinfo {pages} {81} (\bibinfo {year} {1999})}\BibitemShut
  {NoStop}%
\bibitem [{\citenamefont {Meyer}\ \emph {et~al.}(2008)\citenamefont {Meyer},
  \citenamefont {Biedermann}, \citenamefont {Gubo}, \citenamefont {Hammer},\
  and\ \citenamefont {Heinz}}]{Meyer_JPCM_2008}%
  \BibitemOpen
  \bibfield  {author} {\bibinfo {author} {\bibfnamefont {W.}~\bibnamefont
  {Meyer}}, \bibinfo {author} {\bibfnamefont {K.}~\bibnamefont {Biedermann}},
  \bibinfo {author} {\bibfnamefont {M.}~\bibnamefont {Gubo}}, \bibinfo {author}
  {\bibfnamefont {L.}~\bibnamefont {Hammer}}, \ and\ \bibinfo {author}
  {\bibfnamefont {K.}~\bibnamefont {Heinz}},\ }\href {\doibase
  10.1088/0953-8984/20/26/265011} {\bibfield  {journal} {\bibinfo  {journal}
  {J. Phys.: Condens. Matter}\ }\textbf {\bibinfo {volume} {20}},\ \bibinfo
  {pages} {265011} (\bibinfo {year} {2008})}\BibitemShut {NoStop}%
\bibitem [{\citenamefont {Okada}\ \emph {et~al.}(2017)\citenamefont {Okada},
  \citenamefont {Ando}, \citenamefont {Shimizu}, \citenamefont {Minamitani},
  \citenamefont {Shiraki}, \citenamefont {Watanabe},\ and\ \citenamefont
  {Hitosugi}}]{Okada_NatComm_2017}%
  \BibitemOpen
  \bibfield  {author} {\bibinfo {author} {\bibfnamefont {Y.}~\bibnamefont
  {Okada}}, \bibinfo {author} {\bibfnamefont {Y.}~\bibnamefont {Ando}},
  \bibinfo {author} {\bibfnamefont {R.}~\bibnamefont {Shimizu}}, \bibinfo
  {author} {\bibfnamefont {E.}~\bibnamefont {Minamitani}}, \bibinfo {author}
  {\bibfnamefont {S.}~\bibnamefont {Shiraki}}, \bibinfo {author} {\bibfnamefont
  {S.}~\bibnamefont {Watanabe}}, \ and\ \bibinfo {author} {\bibfnamefont
  {T.}~\bibnamefont {Hitosugi}},\ }\href {\doibase 10.1038/ncomms15975}
  {\bibfield  {journal} {\bibinfo  {journal} {Nature Commun.}\ }\textbf
  {\bibinfo {volume} {8}},\ \bibinfo {pages} {15975} (\bibinfo {year}
  {2017})}\BibitemShut {NoStop}%
\bibitem [{\citenamefont {Kitta}\ \emph {et~al.}(2014)\citenamefont {Kitta},
  \citenamefont {Matsuda}, \citenamefont {Maeda}, \citenamefont {Akita},
  \citenamefont {Tanaka}, \citenamefont {Kido},\ and\ \citenamefont
  {Kohyama}}]{Kitta_SS_2014}%
  \BibitemOpen
  \bibfield  {author} {\bibinfo {author} {\bibfnamefont {M.}~\bibnamefont
  {Kitta}}, \bibinfo {author} {\bibfnamefont {T.}~\bibnamefont {Matsuda}},
  \bibinfo {author} {\bibfnamefont {Y.}~\bibnamefont {Maeda}}, \bibinfo
  {author} {\bibfnamefont {T.}~\bibnamefont {Akita}}, \bibinfo {author}
  {\bibfnamefont {S.}~\bibnamefont {Tanaka}}, \bibinfo {author} {\bibfnamefont
  {Y.}~\bibnamefont {Kido}}, \ and\ \bibinfo {author} {\bibfnamefont
  {M.}~\bibnamefont {Kohyama}},\ }\href {\doibase
  https://doi.org/10.1016/j.susc.2013.09.026} {\bibfield  {journal} {\bibinfo
  {journal} {Surf. Sci.}\ }\textbf {\bibinfo {volume} {619}},\ \bibinfo {pages}
  {5} (\bibinfo {year} {2014})}\BibitemShut {NoStop}%
\bibitem [{\citenamefont {Mishra}\ and\ \citenamefont
  {Thomas}(1977)}]{Mishra_JAP_1977}%
  \BibitemOpen
  \bibfield  {author} {\bibinfo {author} {\bibfnamefont {R.~K.}\ \bibnamefont
  {Mishra}}\ and\ \bibinfo {author} {\bibfnamefont {G.}~\bibnamefont
  {Thomas}},\ }\href {\doibase 10.1063/1.323486} {\bibfield  {journal}
  {\bibinfo  {journal} {J. Appl. Phys.}\ }\textbf {\bibinfo {volume} {48}},\
  \bibinfo {pages} {4576} (\bibinfo {year} {1977})}\BibitemShut {NoStop}%
\bibitem [{\citenamefont {Wen}\ \emph {et~al.}(2020)\citenamefont {Wen},
  \citenamefont {Liu}, \citenamefont {Kareev}, \citenamefont {Wu},
  \citenamefont {Terilli}, \citenamefont {Chakhalian}, \citenamefont {Shafer},\
  and\ \citenamefont {Arenholz}}]{Wen_PRB_2020}%
  \BibitemOpen
  \bibfield  {author} {\bibinfo {author} {\bibfnamefont {F.}~\bibnamefont
  {Wen}}, \bibinfo {author} {\bibfnamefont {X.}~\bibnamefont {Liu}}, \bibinfo
  {author} {\bibfnamefont {M.}~\bibnamefont {Kareev}}, \bibinfo {author}
  {\bibfnamefont {T.-C.}\ \bibnamefont {Wu}}, \bibinfo {author} {\bibfnamefont
  {M.}~\bibnamefont {Terilli}}, \bibinfo {author} {\bibfnamefont
  {J.}~\bibnamefont {Chakhalian}}, \bibinfo {author} {\bibfnamefont
  {P.}~\bibnamefont {Shafer}}, \ and\ \bibinfo {author} {\bibfnamefont
  {E.}~\bibnamefont {Arenholz}},\ }\href {\doibase 10.1103/PhysRevB.102.165426}
  {\bibfield  {journal} {\bibinfo  {journal} {Phys. Rev. B}\ }\textbf {\bibinfo
  {volume} {102}},\ \bibinfo {pages} {165426} (\bibinfo {year}
  {2020})}\BibitemShut {NoStop}%
\bibitem [{\citenamefont {Matsushita}, \citenamefont {Ueda},\ and\
  \citenamefont {Ueda}(2005)}]{Matsushita_NatMat_2005}%
  \BibitemOpen
  \bibfield  {author} {\bibinfo {author} {\bibfnamefont {Y.}~\bibnamefont
  {Matsushita}}, \bibinfo {author} {\bibfnamefont {H.}~\bibnamefont {Ueda}}, \
  and\ \bibinfo {author} {\bibfnamefont {Y.}~\bibnamefont {Ueda}},\ }\href
  {\doibase 10.1038/nmat1499} {\bibfield  {journal} {\bibinfo  {journal} {Nat.
  Mater.}\ }\textbf {\bibinfo {volume} {4}},\ \bibinfo {pages} {845} (\bibinfo
  {year} {2005})}\BibitemShut {NoStop}%
\bibitem [{\citenamefont {Ohsawa}\ \emph {et~al.}(2020)\citenamefont {Ohsawa},
  \citenamefont {Yamada}, \citenamefont {Kumatani}, \citenamefont {Takagi},
  \citenamefont {Suzuki}, \citenamefont {Shimizu}, \citenamefont {Shiraki},
  \citenamefont {Nojima},\ and\ \citenamefont {Hitosugi}}]{Ohsawa_ACS_2020}%
  \BibitemOpen
  \bibfield  {author} {\bibinfo {author} {\bibfnamefont {T.}~\bibnamefont
  {Ohsawa}}, \bibinfo {author} {\bibfnamefont {N.}~\bibnamefont {Yamada}},
  \bibinfo {author} {\bibfnamefont {A.}~\bibnamefont {Kumatani}}, \bibinfo
  {author} {\bibfnamefont {Y.}~\bibnamefont {Takagi}}, \bibinfo {author}
  {\bibfnamefont {T.}~\bibnamefont {Suzuki}}, \bibinfo {author} {\bibfnamefont
  {R.}~\bibnamefont {Shimizu}}, \bibinfo {author} {\bibfnamefont
  {S.}~\bibnamefont {Shiraki}}, \bibinfo {author} {\bibfnamefont
  {T.}~\bibnamefont {Nojima}}, \ and\ \bibinfo {author} {\bibfnamefont
  {T.}~\bibnamefont {Hitosugi}},\ }\href {\doibase 10.1021/acsaelm.9b00751}
  {\bibfield  {journal} {\bibinfo  {journal} {ACS Appl. Electron. Mater.}\
  }\textbf {\bibinfo {volume} {2}},\ \bibinfo {pages} {517} (\bibinfo {year}
  {2020})}\BibitemShut {NoStop}%
\bibitem [{\citenamefont {Oura}\ \emph {et~al.}(2003)\citenamefont {Oura},
  \citenamefont {Lifshits}, \citenamefont {Saranin}, \citenamefont {Zotov},\
  and\ \citenamefont {Katayama}}]{Oura_2003}%
  \BibitemOpen
  \bibfield  {author} {\bibinfo {author} {\bibfnamefont {K.}~\bibnamefont
  {Oura}}, \bibinfo {author} {\bibfnamefont {V.}~\bibnamefont {Lifshits}},
  \bibinfo {author} {\bibfnamefont {A.}~\bibnamefont {Saranin}}, \bibinfo
  {author} {\bibfnamefont {A.}~\bibnamefont {Zotov}}, \ and\ \bibinfo {author}
  {\bibfnamefont {M.}~\bibnamefont {Katayama}},\ }\href
  {https://books.google.de/books?id=HBFtQgAACAAJ} {\emph {\bibinfo {title}
  {{Surface Science: An Introduction}}}}\ (\bibinfo  {publisher} {Springer
  Berlin Heidelberg},\ \bibinfo {year} {2003})\BibitemShut {NoStop}%
\bibitem [{\citenamefont {Goswami}\ \emph {et~al.}(2007)\citenamefont
  {Goswami}, \citenamefont {Bhattacharjee}, \citenamefont {Satpati},
  \citenamefont {Roy}, \citenamefont {Satyam},\ and\ \citenamefont
  {Dev}}]{Goswami_SS_2007}%
  \BibitemOpen
  \bibfield  {author} {\bibinfo {author} {\bibfnamefont {D.}~\bibnamefont
  {Goswami}}, \bibinfo {author} {\bibfnamefont {K.}~\bibnamefont
  {Bhattacharjee}}, \bibinfo {author} {\bibfnamefont {B.}~\bibnamefont
  {Satpati}}, \bibinfo {author} {\bibfnamefont {S.}~\bibnamefont {Roy}},
  \bibinfo {author} {\bibfnamefont {P.}~\bibnamefont {Satyam}}, \ and\ \bibinfo
  {author} {\bibfnamefont {B.}~\bibnamefont {Dev}},\ }\href {\doibase
  https://doi.org/10.1016/j.susc.2006.10.026} {\bibfield  {journal} {\bibinfo
  {journal} {Surf. Sci.}\ }\textbf {\bibinfo {volume} {601}},\ \bibinfo {pages}
  {603} (\bibinfo {year} {2007})}\BibitemShut {NoStop}%
\bibitem [{\citenamefont {Yanina}\ and\ \citenamefont {{Barry
  Carter}}(2002)}]{Yanina_SS_2002}%
  \BibitemOpen
  \bibfield  {author} {\bibinfo {author} {\bibfnamefont {S.~V.}\ \bibnamefont
  {Yanina}}\ and\ \bibinfo {author} {\bibfnamefont {C.}~\bibnamefont {{Barry
  Carter}}},\ }\href {\doibase https://doi.org/10.1016/S0039-6028(02)01561-3}
  {\bibfield  {journal} {\bibinfo  {journal} {Surf. Sci.}\ }\textbf {\bibinfo
  {volume} {511}},\ \bibinfo {pages} {133} (\bibinfo {year}
  {2002})}\BibitemShut {NoStop}%
\bibitem [{\citenamefont {Nedelkoski}\ \emph {et~al.}(2017)\citenamefont
  {Nedelkoski}, \citenamefont {Kepaptsoglou}, \citenamefont {Lari},
  \citenamefont {Wen}, \citenamefont {Booth}, \citenamefont {Oberdick},
  \citenamefont {Galindo}, \citenamefont {Ramasse}, \citenamefont {Evans},
  \citenamefont {Majetich},\ and\ \citenamefont
  {Lazarov}}]{Nedelkoski_SRep_2017}%
  \BibitemOpen
  \bibfield  {author} {\bibinfo {author} {\bibfnamefont {Z.}~\bibnamefont
  {Nedelkoski}}, \bibinfo {author} {\bibfnamefont {D.}~\bibnamefont
  {Kepaptsoglou}}, \bibinfo {author} {\bibfnamefont {L.}~\bibnamefont {Lari}},
  \bibinfo {author} {\bibfnamefont {T.}~\bibnamefont {Wen}}, \bibinfo {author}
  {\bibfnamefont {R.~A.}\ \bibnamefont {Booth}}, \bibinfo {author}
  {\bibfnamefont {S.~D.}\ \bibnamefont {Oberdick}}, \bibinfo {author}
  {\bibfnamefont {P.~L.}\ \bibnamefont {Galindo}}, \bibinfo {author}
  {\bibfnamefont {Q.~M.}\ \bibnamefont {Ramasse}}, \bibinfo {author}
  {\bibfnamefont {R.~F.~L.}\ \bibnamefont {Evans}}, \bibinfo {author}
  {\bibfnamefont {S.}~\bibnamefont {Majetich}}, \ and\ \bibinfo {author}
  {\bibfnamefont {V.~K.}\ \bibnamefont {Lazarov}},\ }\href {\doibase
  10.1038/srep45997} {\bibfield  {journal} {\bibinfo  {journal} {Sci. Rep.}\
  }\textbf {\bibinfo {volume} {7}},\ \bibinfo {pages} {45997} (\bibinfo {year}
  {2017})}\BibitemShut {NoStop}%
\bibitem [{\citenamefont {Redinger}\ \emph {et~al.}(2008)\citenamefont
  {Redinger}, \citenamefont {Ricken}, \citenamefont {Kuhn}, \citenamefont
  {R\"atz}, \citenamefont {Voigt}, \citenamefont {Krug},\ and\ \citenamefont
  {Michely}}]{Redinger_PRL_2008}%
  \BibitemOpen
  \bibfield  {author} {\bibinfo {author} {\bibfnamefont {A.}~\bibnamefont
  {Redinger}}, \bibinfo {author} {\bibfnamefont {O.}~\bibnamefont {Ricken}},
  \bibinfo {author} {\bibfnamefont {P.}~\bibnamefont {Kuhn}}, \bibinfo {author}
  {\bibfnamefont {A.}~\bibnamefont {R\"atz}}, \bibinfo {author} {\bibfnamefont
  {A.}~\bibnamefont {Voigt}}, \bibinfo {author} {\bibfnamefont
  {J.}~\bibnamefont {Krug}}, \ and\ \bibinfo {author} {\bibfnamefont
  {T.}~\bibnamefont {Michely}},\ }\href {\doibase
  10.1103/PhysRevLett.100.035506} {\bibfield  {journal} {\bibinfo  {journal}
  {Phys. Rev. Lett.}\ }\textbf {\bibinfo {volume} {100}},\ \bibinfo {pages}
  {035506} (\bibinfo {year} {2008})}\BibitemShut {NoStop}%
\bibitem [{\citenamefont {Eyert}\ \emph {et~al.}(1999)\citenamefont {Eyert},
  \citenamefont {H\"ock}, \citenamefont {Horn}, \citenamefont {Loidl},\ and\
  \citenamefont {Riseborough}}]{Eyert_EPL_1999}%
  \BibitemOpen
  \bibfield  {author} {\bibinfo {author} {\bibfnamefont {V.}~\bibnamefont
  {Eyert}}, \bibinfo {author} {\bibfnamefont {K.-H.}\ \bibnamefont {H\"ock}},
  \bibinfo {author} {\bibfnamefont {S.}~\bibnamefont {Horn}}, \bibinfo {author}
  {\bibfnamefont {A.}~\bibnamefont {Loidl}}, \ and\ \bibinfo {author}
  {\bibfnamefont {P.~S.}\ \bibnamefont {Riseborough}},\ }\href {\doibase
  10.1209/epl/i1999-00330-9} {\bibfield  {journal} {\bibinfo  {journal} {EPL}\
  }\textbf {\bibinfo {volume} {46}},\ \bibinfo {pages} {762} (\bibinfo {year}
  {1999})}\BibitemShut {NoStop}%
\bibitem [{\citenamefont {Tasker}(1979)}]{Tasker_JPC_1979}%
  \BibitemOpen
  \bibfield  {author} {\bibinfo {author} {\bibfnamefont {P.~W.}\ \bibnamefont
  {Tasker}},\ }\href {\doibase 10.1088/0022-3719/12/22/036} {\bibfield
  {journal} {\bibinfo  {journal} {J. Phys. C: Solid State Phys.}\ }\textbf
  {\bibinfo {volume} {12}},\ \bibinfo {pages} {4977} (\bibinfo {year}
  {1979})}\BibitemShut {NoStop}%
\bibitem [{\citenamefont {Chang}\ \emph {et~al.}(2016)\citenamefont {Chang},
  \citenamefont {Hu}, \citenamefont {Klein}, \citenamefont {Liu}, \citenamefont
  {Sutarto}, \citenamefont {Tanaka}, \citenamefont {Cezar}, \citenamefont
  {Brookes}, \citenamefont {Lin}, \citenamefont {Hsieh}, \citenamefont {Chen},
  \citenamefont {Rata},\ and\ \citenamefont {Tjeng}}]{Chang_PRX_2016}%
  \BibitemOpen
  \bibfield  {author} {\bibinfo {author} {\bibfnamefont {C.~F.}\ \bibnamefont
  {Chang}}, \bibinfo {author} {\bibfnamefont {Z.}~\bibnamefont {Hu}}, \bibinfo
  {author} {\bibfnamefont {S.}~\bibnamefont {Klein}}, \bibinfo {author}
  {\bibfnamefont {X.~H.}\ \bibnamefont {Liu}}, \bibinfo {author} {\bibfnamefont
  {R.}~\bibnamefont {Sutarto}}, \bibinfo {author} {\bibfnamefont
  {A.}~\bibnamefont {Tanaka}}, \bibinfo {author} {\bibfnamefont {J.~C.}\
  \bibnamefont {Cezar}}, \bibinfo {author} {\bibfnamefont {N.~B.}\ \bibnamefont
  {Brookes}}, \bibinfo {author} {\bibfnamefont {H.-J.}\ \bibnamefont {Lin}},
  \bibinfo {author} {\bibfnamefont {H.~H.}\ \bibnamefont {Hsieh}}, \bibinfo
  {author} {\bibfnamefont {C.~T.}\ \bibnamefont {Chen}}, \bibinfo {author}
  {\bibfnamefont {A.~D.}\ \bibnamefont {Rata}}, \ and\ \bibinfo {author}
  {\bibfnamefont {L.~H.}\ \bibnamefont {Tjeng}},\ }\href {\doibase
  10.1103/PhysRevX.6.041011} {\bibfield  {journal} {\bibinfo  {journal} {Phys.
  Rev. X}\ }\textbf {\bibinfo {volume} {6}},\ \bibinfo {pages} {041011}
  (\bibinfo {year} {2016})}\BibitemShut {NoStop}%
\bibitem [{\citenamefont {Walls}\ \emph {et~al.}(2016)\citenamefont {Walls},
  \citenamefont {L\"ubben}, \citenamefont {Palot\'as}, \citenamefont
  {Fleischer}, \citenamefont {Walshe},\ and\ \citenamefont
  {Shvets}}]{Walls_PRB_2016}%
  \BibitemOpen
  \bibfield  {author} {\bibinfo {author} {\bibfnamefont {B.}~\bibnamefont
  {Walls}}, \bibinfo {author} {\bibfnamefont {O.}~\bibnamefont {L\"ubben}},
  \bibinfo {author} {\bibfnamefont {K.}~\bibnamefont {Palot\'as}}, \bibinfo
  {author} {\bibfnamefont {K.}~\bibnamefont {Fleischer}}, \bibinfo {author}
  {\bibfnamefont {K.}~\bibnamefont {Walshe}}, \ and\ \bibinfo {author}
  {\bibfnamefont {I.~V.}\ \bibnamefont {Shvets}},\ }\href {\doibase
  10.1103/PhysRevB.94.165424} {\bibfield  {journal} {\bibinfo  {journal} {Phys.
  Rev. B}\ }\textbf {\bibinfo {volume} {94}},\ \bibinfo {pages} {165424}
  (\bibinfo {year} {2016})}\BibitemShut {NoStop}%
\bibitem [{\citenamefont {Azuma}\ \emph {et~al.}(2013)\citenamefont {Azuma},
  \citenamefont {Dover}, \citenamefont {Grinter}, \citenamefont {Grau-Crespo},
  \citenamefont {Almora-Barrios}, \citenamefont {Thornton}, \citenamefont
  {Oda},\ and\ \citenamefont {Tanaka}}]{Azuma_JPCC_2013}%
  \BibitemOpen
  \bibfield  {author} {\bibinfo {author} {\bibfnamefont {K.}~\bibnamefont
  {Azuma}}, \bibinfo {author} {\bibfnamefont {C.}~\bibnamefont {Dover}},
  \bibinfo {author} {\bibfnamefont {D.~C.}\ \bibnamefont {Grinter}}, \bibinfo
  {author} {\bibfnamefont {R.}~\bibnamefont {Grau-Crespo}}, \bibinfo {author}
  {\bibfnamefont {N.}~\bibnamefont {Almora-Barrios}}, \bibinfo {author}
  {\bibfnamefont {G.}~\bibnamefont {Thornton}}, \bibinfo {author}
  {\bibfnamefont {T.}~\bibnamefont {Oda}}, \ and\ \bibinfo {author}
  {\bibfnamefont {S.}~\bibnamefont {Tanaka}},\ }\href {\doibase
  10.1021/jp3119549} {\bibfield  {journal} {\bibinfo  {journal} {J. Phys. Chem.
  C}\ }\textbf {\bibinfo {volume} {117}},\ \bibinfo {pages} {5126} (\bibinfo
  {year} {2013})}\BibitemShut {NoStop}%
\bibitem [{\citenamefont {Kim}, \citenamefont {Aykol},\ and\ \citenamefont
  {Wolverton}(2015)}]{Kim_PRB_2015}%
  \BibitemOpen
  \bibfield  {author} {\bibinfo {author} {\bibfnamefont {S.}~\bibnamefont
  {Kim}}, \bibinfo {author} {\bibfnamefont {M.}~\bibnamefont {Aykol}}, \ and\
  \bibinfo {author} {\bibfnamefont {C.}~\bibnamefont {Wolverton}},\ }\href
  {\doibase 10.1103/PhysRevB.92.115411} {\bibfield  {journal} {\bibinfo
  {journal} {Phys. Rev. B}\ }\textbf {\bibinfo {volume} {92}},\ \bibinfo
  {pages} {115411} (\bibinfo {year} {2015})}\BibitemShut {NoStop}%
\bibitem [{Spr()}]{Springer2}%
  \BibitemOpen
  \href {https://materials.springer.com} {\emph {\bibinfo {title}
  {{SpringerMaterials database}}}}\BibitemShut {NoStop}%
\bibitem [{\citenamefont {Fujita}\ \emph {et~al.}(1997)\citenamefont {Fujita},
  \citenamefont {Amemiya}, \citenamefont {Yakabe}, \citenamefont {Nejoh},
  \citenamefont {Sato},\ and\ \citenamefont {Iwatsuki}}]{Fujita_PRL_1997}%
  \BibitemOpen
  \bibfield  {author} {\bibinfo {author} {\bibfnamefont {D.}~\bibnamefont
  {Fujita}}, \bibinfo {author} {\bibfnamefont {K.}~\bibnamefont {Amemiya}},
  \bibinfo {author} {\bibfnamefont {T.}~\bibnamefont {Yakabe}}, \bibinfo
  {author} {\bibfnamefont {H.}~\bibnamefont {Nejoh}}, \bibinfo {author}
  {\bibfnamefont {T.}~\bibnamefont {Sato}}, \ and\ \bibinfo {author}
  {\bibfnamefont {M.}~\bibnamefont {Iwatsuki}},\ }\href {\doibase
  10.1103/PhysRevLett.78.3904} {\bibfield  {journal} {\bibinfo  {journal}
  {Phys. Rev. Lett.}\ }\textbf {\bibinfo {volume} {78}},\ \bibinfo {pages}
  {3904} (\bibinfo {year} {1997})}\BibitemShut {NoStop}%
\bibitem [{\citenamefont {{de Picciotto}}\ and\ \citenamefont
  {Thackeray}(1985)}]{DePicciotto_MRB_1985}%
  \BibitemOpen
  \bibfield  {author} {\bibinfo {author} {\bibfnamefont {L.}~\bibnamefont {{de
  Picciotto}}}\ and\ \bibinfo {author} {\bibfnamefont {M.}~\bibnamefont
  {Thackeray}},\ }\href {\doibase https://doi.org/10.1016/0025-5408(85)90158-8}
  {\bibfield  {journal} {\bibinfo  {journal} {Mater. Res. Bull.}\ }\textbf
  {\bibinfo {volume} {20}},\ \bibinfo {pages} {1409} (\bibinfo {year}
  {1985})}\BibitemShut {NoStop}%
\end{thebibliography}

%

\newpage
\onecolumngrid

\setcounter{figure}{0}
\setcounter{equation}{0}
\setcounter{table}{0}
\setcounter{section}{0}
\setcounter{subsection}{0}
\makeatletter
\renewcommand{\thefigure}{S\@arabic\c@figure}
\renewcommand{\theequation}{S\@arabic\c@equation}
\renewcommand{\thetable}{S\@arabic\c@table}
\newcounter{SIfig}
\renewcommand{\theSIfig}{S\arabic{SIfig}}
\newcounter{SIeq}
\renewcommand{\theSIeq}{S\arabic{SIeq}}

\section*{Supplementary Material}

\subsection*{Growth parameters and surface preparation}

The growth parameters and surface preparation of the LiV$_2$O$_4$(111) films reported in this work are given in Tables \ref{TSamples} and \ref{TSurface}.

\begin{table}[h]
\setlength{\tabcolsep}{7pt}
\caption{Film growth parameters.}
\begin{tabular}{cccccccc}
\hline \hline
Sample & Substrate & Deposition & Laser energy & Laser fluence & Pulses & Repetition & Figures \\
& & temperature ($^{\circ}$C) & at output (mJ) & (J/cm$^2$) & & rate (Hz) \\
\hline
A & SrTiO$_3$(111) & 660 & 720 & $\sim$1.5 & 1500 & 5 & 1(b)--1(e), 2(b), \ref{FigS4}(a), \ref{FigS5}  \\ 
B & SrTiO$_3$(111) & 660 & 790 & $\sim$1.7 & 1500 & 5 & 1(f), \ref{FigS4}(b), \ref{FigS4}(c) \\ 
C & SrTiO$_3$(111) & 660 & 720 & $\sim$1.5 & 1500 & 5 & 2(c), 3(a), 3(d), 3(g), 4(a), 4(b) \\ 
&  &  &  &  & &  & \ref{FigS4}(d)--\ref{FigS4}(f), \ref{FigS5} \\ 
D & Nb-SrTiO$_3$(111) & 520 & 780 & $\sim$1.6 & 1500 & 5 & 2(d), 3(b), 3(e), 4(c), \ref{FigS5} \\ 
E & Nb-SrTiO$_3$(111) & 580 & 800 & $\sim$1.7 & 1500 & 5 & 2(e)--2(h), 3(c), 3(f), 3(h), \ref{FigS5} \\ 
F & Nb-SrTiO$_3$(111) & 520 & 780 & $\sim$1.6 & 1000 & 5 & \ref{FigS1} \\ 
G & SrTiO$_3$(111) & 660 & 720 & $\sim$1.5 & 1000 & 5 & \ref{FigS5} \\ 
H & SrTiO$_3$(111) & 660 & 810 & $\sim$1.7 & 1500 & 5 & \ref{FigS5} \\ 
I & Nb-SrTiO$_3$(111) & 580 & 800 & $\sim$1.7 & 1500 & 5 & \ref{FigS5} \\ 
\hline \hline
\end{tabular}
\label{TSamples}
\end{table}

\begin{table}[h]
\setlength{\tabcolsep}{7pt}
\caption{Description of surface preparation.}
\begin{tabular}{c|l}
\hline \hline
Sample & Procedure \\
\hline
A & Anneal up to 530$^{\circ}$C \\ 
C & Sputter up to 750 V in 10$^{-3}$ mbar Ar partial pressure; anneal up to 550$^{\circ}$C \\ 
D & Anneal up to 600$^{\circ}$C \\ 
E & Anneal up to 600$^{\circ}$C \\ 
F & Sputter up to 750 V in 10$^{-3}$ mbar Ar partial pressure; anneal up to 550$^{\circ}$C \\ 
\hline \hline
\end{tabular}
\label{TSurface}
\end{table}

\newpage \subsection*{Additional sample}

\begin{figure}[h]
\includegraphics{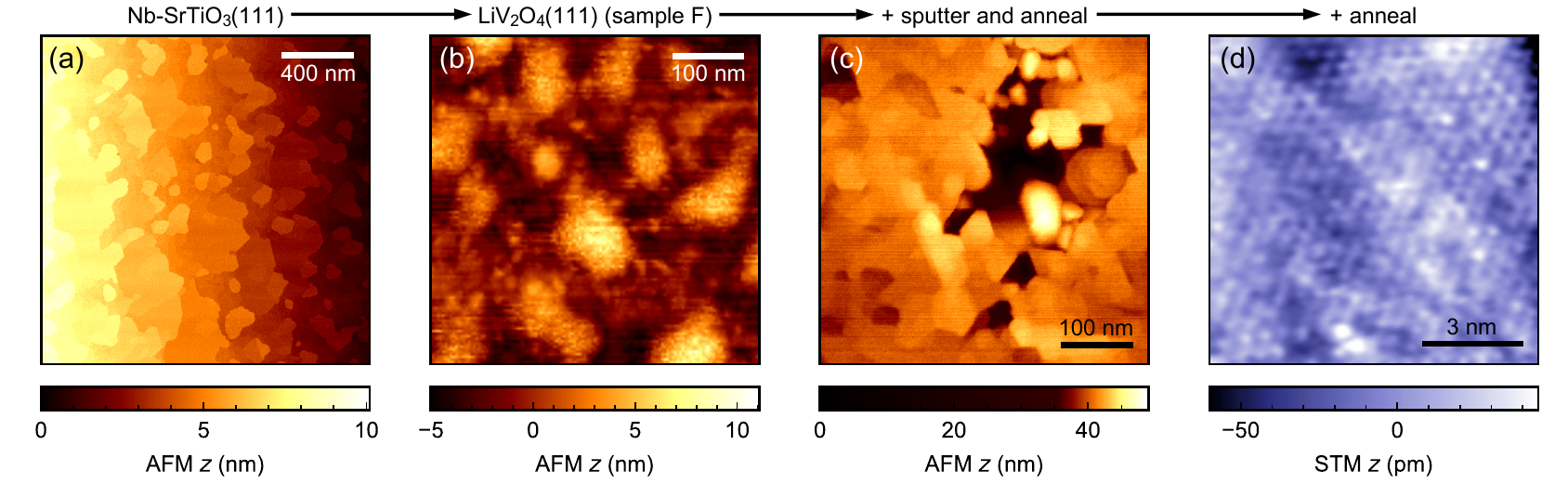}
\caption{\label{FigS1} Topographic images of sample F at various stages of preparation. (a) \textit{Ex-situ} AFM topographic image of starting Nb-SrTiO$_3$(111) substrate after annealing in air at 1000$^{\circ}$C for several hours. (b) \textit{Ex-situ} AFM topographic image of LiV$_2$O$_4$(111) film after PLD. (c) \textit{Ex-situ} AFM topographic image of LiV$_2$O$_4$(111) film after additional sputter and annealing cycles. Sputtering: 750 V, 10$^{-3}$ mbar Ar partial pressure. Annealing: 550$^{\circ}$C. (d) Subsequent \textit{in-situ} STM topographic image of the LiV$_2$O$_4$(111) film. Additional annealing was required after air exposure to clean the surface. Set point: 50 mV, 500 pA. In (d), Gaussian smoothing with a width of one pixel has been applied to enhance the signal-to-noise ratio.}
\end{figure}

Figure \ref{FigS1} presents AFM and STM images of an additional LiV$_2$O$_4$(111) film (sample F) at four different stages: (a) bare substrate, (b) after PLD, (c) after sputtering and annealing, and (d) after additional annealing. As mentioned in the main text, we pretreated the SrTiO$_3$(111) substrates by heating them up to 1000$^{\circ}$C in atmosphere for several hours, in order to obtain flat terraces [Fig.~\ref{FigS1}(a)]. The LiV$_2$O$_4$ film after PLD has a granular surface morphology, as seen in Fig.~\ref{FigS1}(b). After sputtering and annealing, as previously described, AFM reveals triangular, hexagonal and round islands with flat tops [Fig.~\ref{FigS1}(c)]. STM resolves a hexagonal atomic lattice [Fig.~\ref{FigS1}(d)]. 

\newpage \subsection*{Growth window}

\begin{figure}[h]
\includegraphics{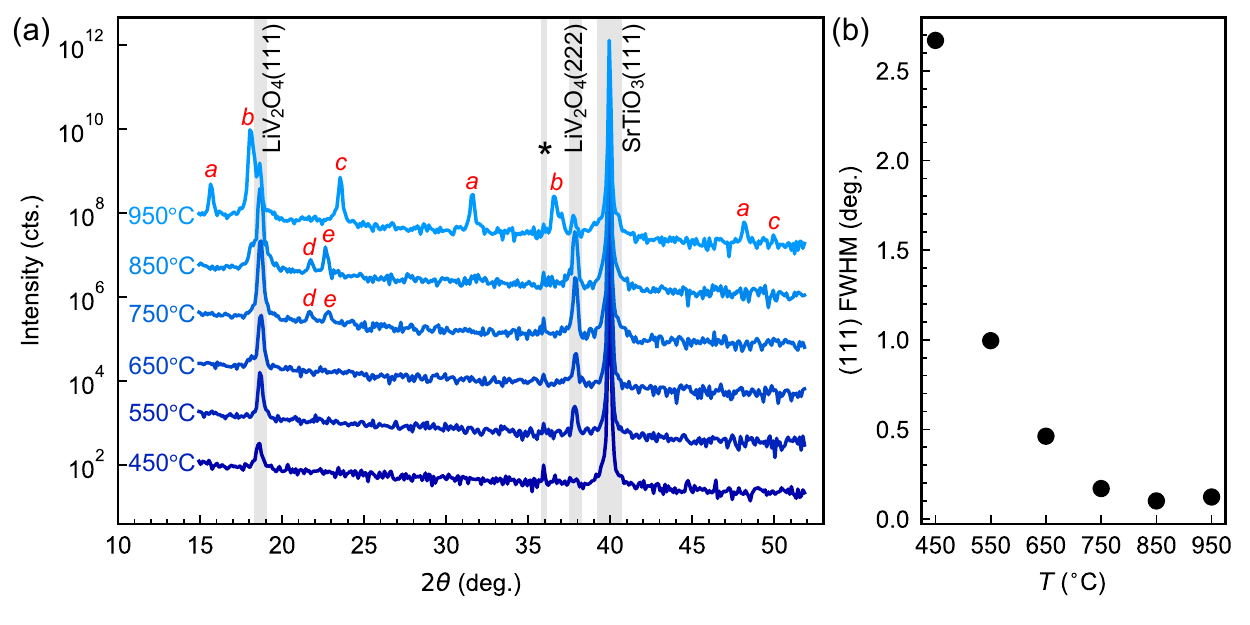}
\caption{\label{FigS2} (a) X-ray $\omega$-2$\theta$ scans for a series of LiV$_2$O$_4$ films deposited on Nb-SrTiO$_3$(111) at various substrate temperatures between 450 and 950$^{\circ}$C. The films are roughly 45--70 nm thick. The red letters, $a$--$e$, mark visible impurity peaks (refer to Table \ref{TImp}). The asterisks mark a spurious reflection of the SrTiO$_3$(111) peak due to an unfiltered Cu K$_{\beta}$ component of the Cu K$_{\alpha}$ source. (b) FWHM of the rocking curve of the LiV$_2$O$_4$(111) peaks as a function of substrate temperature.}
\end{figure}

We investigated the deposition of LiV$_2$O$_4$(111) films at substrate temperatures ranging from 450 to 950$^{\circ}$C. As seen in Fig.~\ref{FigS2}(a), at the lowest temperature of 450$^{\circ}$C, the amplitude of the LiV$_2$O$_4$(111) Bragg reflection in XRD is weak, and the LiV$_2$O$_4$(222) reflection is not visible. As we increased the temperature, the amplitudes of both reflections increased and reached a maximum at 750--850$^{\circ}$C. However, impurity peaks are also visible starting around 750$^{\circ}$C. At 950$^{\circ}$C, it is clear that LiV$_2$O$_4$ is no longer the dominant phase in the film. The impurity phases are identified as various vanadates in Table \ref{TImp}, which suggests a loss of Li at temperatures above 750$^{\circ}$C.    

According to Fig.~\ref{FigS2}(b), the LiV$_2$O$_4$ phase has the best crystallinity at 850$^{\circ}$C, with the lowest FWHM of the (111) Bragg reflection ($\approx$0.1$^{\circ}$). However, the film is no longer phase pure. For the measurements presented in the main text, we have restricted the growth and vacuum annealing temperatures to the safer ranges of 520--660$^{\circ}$C and 500--600$^{\circ}$C, respectively. The slight difference in temperature ranges has to do with the temperature calibration of different systems, as well as some differences in heating of insulating SrTiO$_3$ and conducting Nb-SrTiO$_3$.   

\begin{table}[h]
\setlength{\tabcolsep}{7pt}
\caption{Impurity phases in Fig.~\ref{FigS2}(a). The dataset ID correspond to those listed in the SpringerMaterials database \cite{Springer2}. For impurity peak $e$, we were unable to identify a candidate compound.}
\begin{tabular}{cc|ccc}
\hline \hline
Label & 2$\theta$ ($^{\circ}$) & Possible identity of reflections & 2$\theta$ ($^{\circ}$) & Dataset ID \\
\hline
$a$ & 15.67, 31.59, 48.17 & V$_2$O$_5$(020), V$_2$O$_5$(040), V$_2$O$_5$(060) & 15.37, 31.03, 47.30 & sd-1500170  \\ 
 &  & VO$_2$(B)(200), VO$_2$(B)(400), VO$_2$(B)(600) & 15.36, 31.00, 47.27 & sd-1614307  \\ 
$b$ & 18.08, 36.59 & VO$_2$(100), VO$_2$(200) & 18.29, 37.06 & sd-0455502 \\ 
$c$ & 23.55, 49.93 & VO$_2$(B)(201), VO$_2$(B)(402) & 23.96, 49.06 & sd-1614307 \\ 
$d$ & 21.67 & V$_3$O$_7$(204) & 21.78 & sd-0308766 \\ 
 &  & V$_3$O$_5$(011) & 22.18 & sd-1724204 \\ 
 $e$ & 22.80 & -- & -- & -- \\ 
\hline \hline
\end{tabular}
\label{TImp}
\end{table}

\newpage \subsection*{Calibration of STM $z$ height}

\begin{figure}[h]
\includegraphics{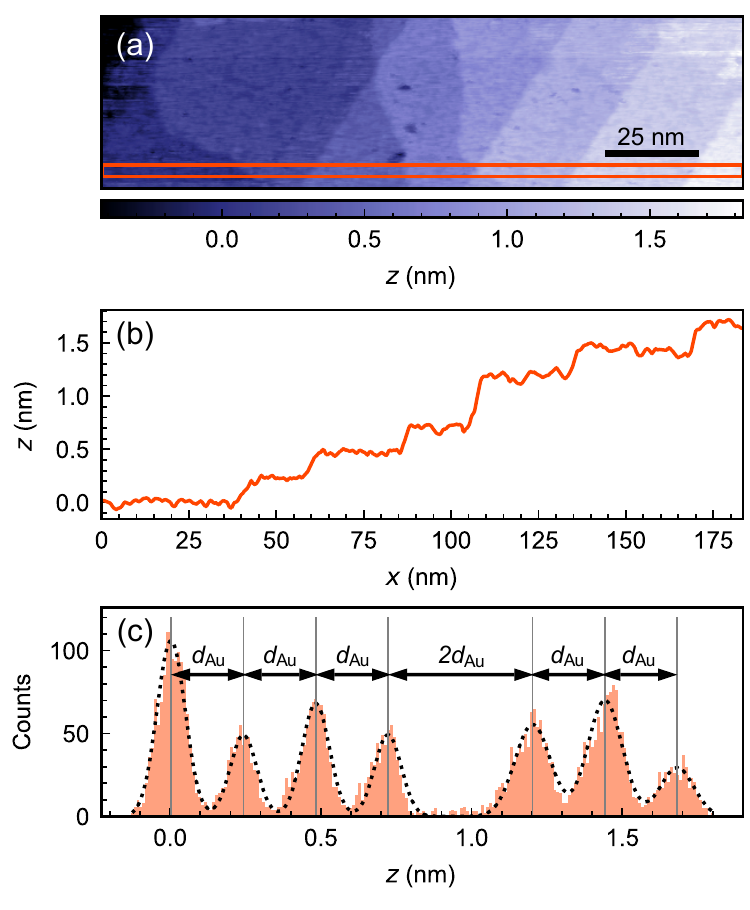}
\caption{\label{FigS3} Calibration of STM scanner in the $z$ direction. (a) Topographic image of Au(111) with repeating terraces. Setpoint: 1 V, 100 pA. (b) Row-averaged height profile within the orange box in (a). (c) Histogram computed within the orange box in (a). The dashed line represents a fit to a sum of Gaussian functions. The $z$ height is scaled such that $d_{\textrm{Au}} \approx 0.24$ nm.}
\end{figure}

To ensure that the observation of irregular step heights in LiV$_2$O$_4$ films in Fig.~3 of the main text does not arise from an instrument artifact, we present control measurements of Au(111) with regular step heights in Fig.~\ref{FigS3}. From the orange box in Fig.~\ref{FigS3}(a) spanning several Au terraces, we computed a $z$-height histogram [Fig.~\ref{FigS3}(c)], which revealed peaks at regularly spaced intervals. We could fit the histogram with a sum of Gaussian functions whose peaks are spaced by $d_{\textrm{Au}}$ or 2$d_{\textrm{Au}}$; i.e., the terrace heights are integer multiples of a single variable, $d_{\textrm{Au}}$. We then calibrated the STM $z$ height to match $d_{\textrm{Au}}$ with its literature value of 0.24 nm \cite{Fujita_PRL_1997}.

\newpage \subsection*{Bulk properties after sputtering and annealing}

\begin{figure}[h]
\includegraphics{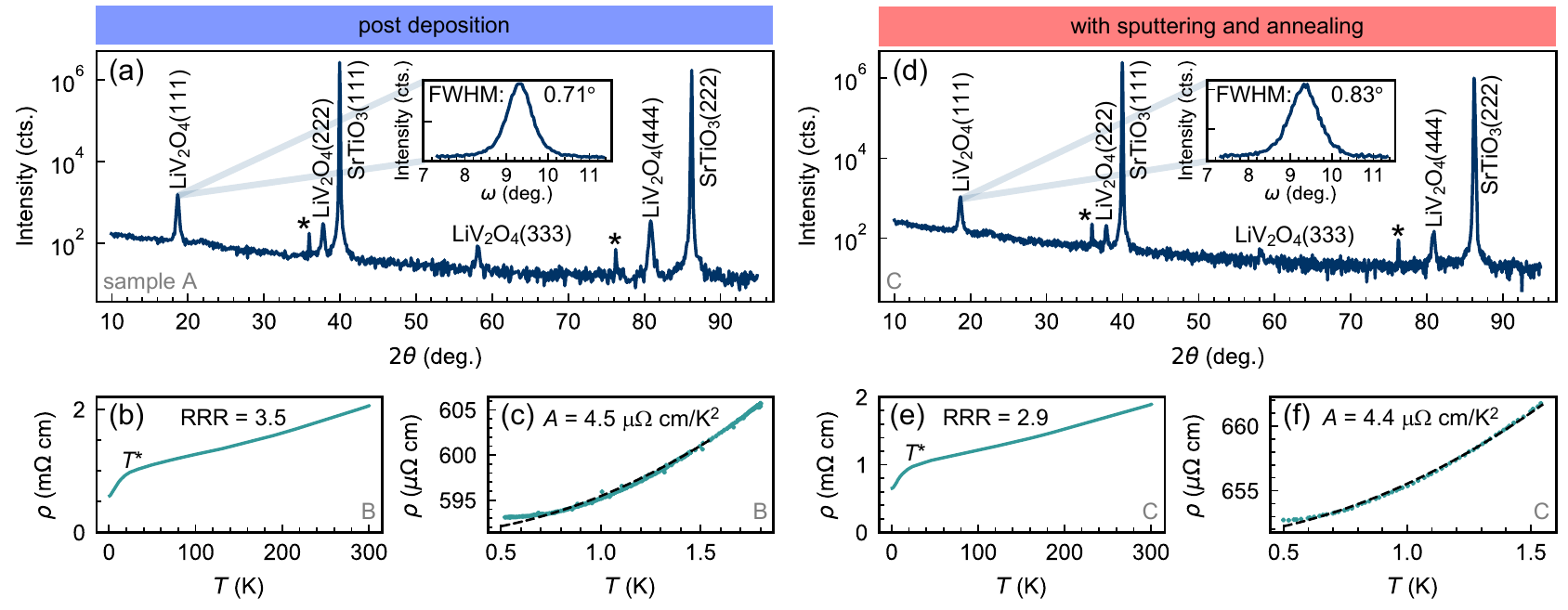}
\caption{\label{FigS4} Comparison of XRD and transport of LiV$_2$O$_4$ films (a)--(c) after PLD (samples A and B) and (d)--(f) after additional sputtering and annealing (sample C). All films were deposited on insulating SrTiO$_3$(111) at 660$^{\circ}$C. Panel (a) is reproduced from Fig.~1(c), whereas panels (b) and (c) are reproduced from Fig.~1(f). In (a) and (d), the asterisks mark spurious reflections of the dominant substrate peaks SrTiO$_3$(111) and SrTiO$_3$(222) due to an unfiltered Cu K$_{\beta}$ component of the Cu K$_{\alpha}$ source, and the insets show rocking curves of the LiV$_2$O$_4$(111) peak, with FWHM of 0.71$^{\circ}$ and 0.83$^{\circ}$. The extracted lattice constants are $a$ = 8.238 and 8.299~\AA, respectively. In (b) and (c), the residual resistivity ratio (RRR) is 3.5 and the quadratic coefficient of the temperature dependence between 0.5 and 1.8 K is $A$ = 4.5 $\mu \Omega$ cm/K$^2$. In (e) and (f), the RRR is 2.9 and $A$ is 4.4 $\mu \Omega$ cm/K$^2$ between 0.5 and 1.5 K.}
\end{figure}

Figure \ref{FigS4} presents a comparison of XRD and transport of LiV$_2$O$_4$ films deposited on insulating SrTiO$_3$(111) at 660$^{\circ}$C, immediately after PLD [Figs.~\ref{FigS4}(a)--\ref{FigS4}(c)] and after sputtering and annealing to produce surfaces suitable for STM [Figs.~\ref{FigS4}(d)--\ref{FigS4}(f)]. For the sputtered and annealed film (sample C), we observed the same Bragg reflections associated with the (111) planes of LiV$_2$O$_4$ [Fig.~\ref{FigS4}(d)]. The resistivity exhibits metallic behavior with a downturn at $T^* \approx 20$ K and $T^2$-like behavior below 1.5 K [Figs.~\ref{FigS4}(e)--\ref{FigS4}(f)]. The reduced amplitude of the Bragg reflections, larger full-width half-maximum of the LiV$_2$O$_4$(111) rocking curve, reduced residual resistivity ratio, and larger lattice constant $a$ (Fig.~\ref{FigS5}) could reflect some effects of the sputtering and annealing, but these differences lie within the variation of our samples. 

\newpage \subsection*{Estimation of Li stoichiometry from lattice constant}

\begin{figure}[h]
\includegraphics{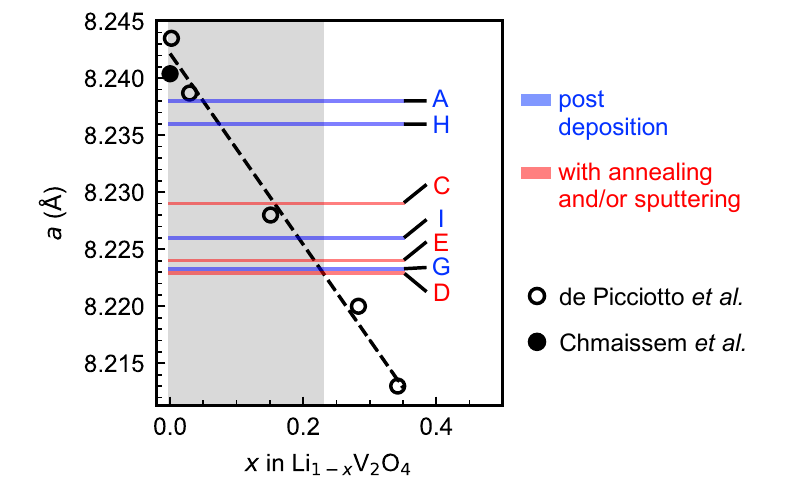}
\caption{\label{FigS5} Cubic lattice constant $a$ of Li$_{1-x}$V$_2$O$_4$ vs. concentration of Li deficiencies $x$. The circles represent data points reproduced from bulk studies (open circles \cite{DePicciotto_MRB_1985}; filled circle \cite{Chmaissem_PRL_1997}), and the dashed line is a linear fit to the data from Ref.\cite{DePicciotto_MRB_1985}. The solid horizontal lines denote the $a$ values measured for various LiV$_2$O$_4$ thin films grown throughout this work, either immediately after deposition (blue), or after an additional annealing and/or sputtering step (red). The capital letters denote the sample names listed in Table \ref{TSamples}, where growth conditions are given. The intersections of the horizontal lines with the dashed line yield estimates of $x$ in the films.}
\end{figure}

Due to the small atomic mass of Li and the thinness of our films, we were not able to accurately quantify the Li concentration with state-of-the-art energy-dispersive x-ray spectroscopy (EDX) instruments available to us. Indirectly, there is an empirical correlation between the Li deficiency $x$ in Li$_{1-x}$V$_2$O$_4$ and the cubic lattice constant $a$, based on powder sample studies \cite{DePicciotto_MRB_1985}. As seen in Fig.~\ref{FigS5}, we extract the $a$ values for several films based on the ($lll$) Bragg reflections and overlay them on the trend line of Li$_{1-x}$V$_2$O$_4$. The films were grown throughout the course of our investigation using various growth and surface treatment conditions documented in Tables \ref{TSamples} and \ref{TSurface}. We roughly estimate the Li deficiency in our films to be $x < 0.23$ (gray shaded area in Fig.~\ref{FigS5}). 

\newpage \subsection*{Volmer-Weber growth of LiV$_2$O$_4$ on SrTiO$_3$(111)}

From STM images of $\sim$50-nm-thick films (Fig.~3 of the main text), we cannot distinguish whether LiV$_2$O$_4$ grows on SrTiO$_3$(111) in Volmer-Weber (VW) or Stranski-Krastanov (SK) mode. In the former, the deposited film forms islands starting at the substrate interface, whereas in the latter, the deposited film initially grows layer-by-layer at the substrate interface, before reverting to island growth past a critical thickness~\cite{Oura_2003}. SK growth mode occurs when there is strong bonding between the film and substrate atoms. 

\begin{figure}[h]
\includegraphics{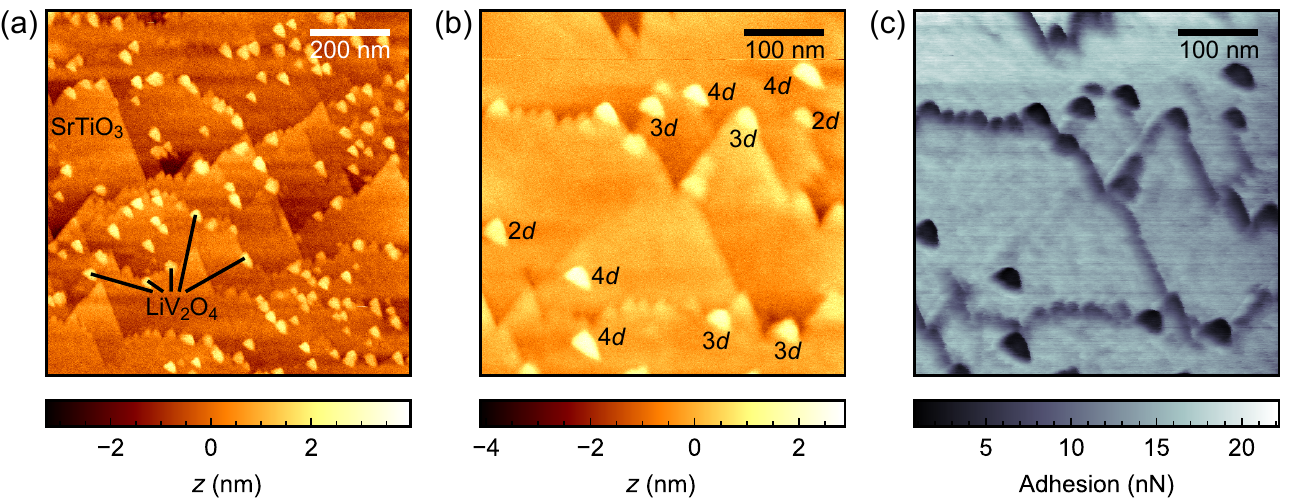}
\caption{\label{FigS6} (a) and (b) AFM topographic images of LiV$_2$O$_4$ islands on SrTiO$_3$(111). In (b), the average heights of ten islands are labeled in units of $d$, the length of the unit cell along the [111] direction. (c) AFM adhesion map acquired simultaneously with (b).}
\end{figure}

To investigate the initial stages of growth, we deposited 15 pulses of LiV$_2$O$_4$ onto a SrTiO$_3$(111) substrate held at 370$^{\circ}$C. (A different PLD chamber was used in this case, leading to slightly adjusted parameters.) Figure~\ref{FigS6}(a) shows a topography acquired \textit{ex situ} via AFM. Rather than exhibiting a uniform coverage, the film is clustered into triangular islands, indicating that the film adatoms were mobile and underwent significant diffusion at the given deposition conditions. Some islands are lodged at the step edges of the SrTiO$_3$(111) substrate, but many others are also located in the middle of the terraces. The triangular islands are identically oriented, which agrees with the RHEED data (Fig.~2 of the main text) showing the thick films to have clear in-plane orientation. We conclude that LiV$_2$O$_4$ on SrTiO$_3$(111) exhibits VW, not SK, growth mode. 

Figure~\ref{FigS6}(b) shows a smaller 500$\times$500 nm field of view. The average heights of the islands range from 1.0 to 2.2 nm, corresponding roughly to 2--4$d$. For the ten islands labeled in Fig.~\ref{FigS6}(b), the lateral areas range from 500 to 1100 nm$^2$, which are comparable to the island sizes observed in Fig.~3 of the main text. Unlike the thick films, however, only triangular, not hexagonal, islands are observed here. It is possible that the longer deposition time of thicker films, or the vacuum annealing procedure preceding STM imaging of those films, promoted further adatom diffusion and compactification of triangular islands into hexagonal islands, which are thermodynamically more stable~\cite{Oura_2003}.

Figure~\ref{FigS6}(c) shows an AFM adhesion map acquired simultaneously with topography in Fig.~\ref{FigS6}(b). The map shows chemical contrast in the form of a lower adhesion for LiV$_2$O$_4$ compared to SrTiO$_3$. However, the measurement is also sensitive to geometry, as the SrTiO$_3$ terrace edges also exhibit reduced adhesion compared to the rest of the terrace.

\end{document}